# Core-mantle partitioning and the bulk Earth abundances of hydrogen and carbon: Implications for their origins

Yutaro Tsutsumi[1], Naoya Sakamoto[2], Kei Hirose[1,3]*, Shuhei Mita[1], Shunpei Yokoo[1], Han Hsu[4] & Hisayoshi Yurimoto[2,5]

[1] Department of Earth and Planetary Science, The University of Tokyo, Bunkyo, Tokyo 113-0033, Japan

[2] Institute for Integrated Innovations, Hokkaido University, Sapporo, Hokkaido 001-0021, Japan

[3] Earth-Life Science Institute, Tokyo Institute of Technology, Meguro, Tokyo 152-8550, Japan

[4] Department of Physics, National Central University, Taoyuan City 320317, Taiwan

[5] Department of Natural History Sciences, Hokkaido University, Sapporo, Hokkaido 060-0810, Japan

*email: kei@eps.s.u-tokyo.ac.jp

We determined the metal/silicate partition coefficients of hydrogen and carbon, $D_H$ and $D_C$, simultaneously under typical conditions of Earth's core formation. Experiments demonstrate that both $D_H$ and $D_C$ diminish in the presence of carbon and hydrogen, respectively, indicating their strong interactions in liquid metal. With these partitioning data, we investigated the core and bulk Earth abundances of hydrogen and carbon based on core formation scenarios that are compatible with the bulk silicate Earth composition and the mass fraction and density deficit of the core. The results of the single-stage core formation modelling are markedly different from those using $D_H$ and $D_C$ individually determined in earlier experiments, indicating that the Earth building blocks do not match enstatite chondrites in water abundance and require contributions by carbonaceous chondrites. The multi-stage core formation models combined with an Earth accretion scenario accounting for isotopic composition show 0.18–0.49 wt% H and 0.19–1.37 wt% C in the core, leading to 0.53–1.40 wt% $H_2O$ (present as H in the core) and 0.07–0.44 wt% C in the bulk Earth. Our modelling also demonstrates that up to 53% and 72% of Earth's water (hydrogen) and carbon, respectively, could have been derived from non-carbonaceous chondritic materials.

--------------------------------------------------------------------------------------

Hydrogen and carbon are important volatile elements on our planet, and their sources[1,2] and the timing[3] of the delivery to the growing Earth are of great interest. Since both hydrogen[4] and carbon[5,6] are known to be strongly siderophile (iron-loving) under high pressure and temperature (*P-T*), the core is likely to be their largest reservoir. Indeed, the Earth's outer core exhibits a large density deficit with respect to pure iron[7], indicating the presence of substantial amounts of light impurity elements, but core concentration of each light element has been controversial[8]. In addition to their bulk silicate Earth (BSE) abundances[9,10], understanding the hydrogen and carbon inventories of the core is essential to estimate their bulk Earth contents, which may suggest their origins.

The hydrogen and carbon contents in the core can be constrained by a combination of their metal/silicate partition coefficients under high *P-T* conditions of core formation and their BSE abundances including the ocean water. The metal-silicate partitioning of hydrogen has been examined by experiments[4] to 60 GPa and 4560 K as well as by theoretical calculations[11,12]. The partitioning of carbon has been extensively studied below 25 GPa in large-volume presses[13-17] and to 71 GPa in pressure and 5200 K in temperature by laser-heated diamond-anvil cell (DAC) techniques[5,6]. However, these two earlier DAC works reported apparently inconsistent results although experiments were conducted at similar *P-T* conditions.

Here we examined the metal-silicate partitioning of hydrogen and carbon simultaneously in a DAC based on a combination of synchrotron X-ray diffraction (XRD) measurements at high pressure and the textural and compositional analyses on recovered samples by secondary ion mass spectrometry (SIMS) and field-emission-type electron probe microanalyzer (FE-EPMA). With these partitioning data, we modelled the Earth's accretion and core formation processes that account for the BSE composition and the mass and density deficit of the core, which give core concentrations of hydrogen and carbon and accordingly their bulk Earth abundances. These results suggest that the core is a predominant reservoir for both hydrogen and carbon on Earth. We also found that non-carbonaceous chondritic materials could be the important source of the Earth's water and carbon.

**Results**

Ten separate melting experiments were performed at 33–56 GPa and 3630–4760 K, corresponding to typical conditions of Earth's core formation[18-20] under both hydrous and anhydrous conditions (Table 1). After melting at high pressures, all samples were recovered from a DAC and examined for melting texture and chemical composition on their cross sections at the centre of a heated area (Fig. 1). Quenched liquid metal was found at the central portion, being surrounded by silicate melt. The davemaoite (CaSiO$_3$-perovskite) and SiO$_2$ layers were present outside of the silicate melt pool, which is consistent with the XRD observations during melting (Fig. 2). The further outside, colder region was under subsolidus temperatures.

**Partitioning of hydrogen.** The metal/silicate partition coefficients of hydrogen, $D_\text{H}^\text{metal/silicate}$ were obtained as:

$$D_\text{H}^\text{metal/silicate} = \frac{x'^\text{metal}_\text{H}}{x^\text{silicate}_{\text{HO}_{0.5}}} \quad (1)$$

where $x'^\text{metal}_\text{H}$ and $x^\text{silicate}_{\text{HO}_{0.5}}$ represent the molar fractions of hydrogen in metal and silicate, respectively (see Eq. 10 for $x'^\text{metal}_i$ in "Methods"). The $D_\text{H}$ values obtained in this study in the presence of carbon ranged from 0.7 to 26.1 (Table 1), notably lower than the $D_\text{H}^\text{metal/silicate}$ = 22–46 measured in a carbon-poor system by previous DAC experiments[4].

The metal-silicate partitioning of hydrogen can be expressed as a chemical reaction:

$$\text{HO}_{0.5}^\text{silicate melt} + \frac{1}{2}\text{Fe}^\text{metal} = \text{H}^\text{metal} + \frac{1}{2}\text{FeO}^\text{silicate melt} \quad (2)$$

The exchange coefficient $K_D$ for this reaction is parameterized as functions of $P$ and $T$ with regression constants $a$, $b$, $c$ and $d$:

$$\log_{10} K_D = \log_{10} \frac{x'^\text{metal}_\text{H}}{x^\text{silicate}_{\text{HO}_{0.5}}} \cdot \sqrt{\frac{x^\text{silicate}_\text{FeO}}{x'^\text{metal}_\text{Fe}}} = \log_{10} \frac{x'^\text{metal}_\text{H}}{x^\text{silicate}_{\text{HO}_{0.5}}} + \frac{1}{4}\Delta\text{IW}$$

$$= \log_{10} D_\text{H}^\text{metal/silicate} + \frac{1}{4}\Delta\text{IW} = a + \frac{b}{T} + c \cdot \frac{P}{T} + d \cdot \log_{10}(1 - x'^\text{metal}_\text{C}) \quad (3)$$

in which $x'^\text{metal}_\text{C}$ is the molar fraction of carbon in liquid metal (Eq. 10), and $d \cdot \log_{10}(1 - x'^\text{metal}_\text{C})$ approximates a non-ideal interaction between H and C[5,21]. Following previous studies[4,22], oxygen fugacity $fO_2$ relative to the iron-wüstite (IW) buffer is approximated as $\Delta\text{IW} \approx 2\log_{10}(x^\text{silicate}_\text{FeO}/x'^\text{metal}_\text{Fe})$, which was obtained from the EPMA analyses. Fitting Eq. 3 to the present and earlier data[4] yields $a$ = 2.31(20), $b$ = -1542(500), $c$ = -38.9(78) and $d$ = 6.69(30). Fig. 3a shows the negative pressure and positive temperature dependence of $K_D$ for the partitioning of hydrogen after correcting for the effect of carbon, consistent with the results of Tagawa et al.[4] We note that carbon in liquid metal strongly reduces $D_\text{H}^\text{metal/silicate}$ (Fig. 3b). All of the experimental data obtained in this study and Tagawa et al.[4] are well reproduced by Eq. 3 (Supplementary Fig. 1a).

**Partitioning of carbon.** Similarly, the metal/silicate partition coefficients for carbon $D_\text{C}^\text{metal/silicate}$ were also calculated as:

$$D_C^{\text{metal/silicate}} = \frac{x'^{\text{metal}}_C}{x^{\text{silicate}}_C} \qquad (4)$$

The $D_C$ ranged from 14 to 369 in the hydrogen (water)-free system and from 27 to 132 in hydrogen (water)-bearing system (Table 1). We fitted the following equation to the present results combined with earlier DAC data[5,6] and low $P$-$T$ data obtained for silicate melts with $nbo/t$ (<10) from multi-anvil and piston-cylinder experiments[13-15,17] that were performed in carbon-undersaturated systems:

$$\log_{10} D_C = a + b \cdot \Delta\text{IW} + c \cdot nbo/t + d \cdot \log_{10}(1 - x'^{\text{metal}}_H) \qquad (5)$$

The fitting provides $a$ = 1.51(30), $b$ = -0.68(15), $c$ = -0.33(7) and $d$ = 1.76(77), showing the strong negative effects of oxygen fugacity, $nbo/t$ and hydrogen concentration in liquid metal on $D_C$ (see Figs. 4a–c, respectively). Note that metal compositions were calculated as $x'^{\text{metal}}_i$ also for those previous experimental data.

Considering the effects of oxygen fugacity and $nbo/t$, all of these earlier experimental data are well reproduced by Eq. 5 (Supplementary Fig. 1b). The $P$ and $T$ dependence of $D_C$ is not explicit in Eq. 5 unlike previous DAC studies[5,6] (Supplementary Fig. 2). The low $P$-$T$ data used together with the high $P$-$T$ DAC data are key to clarifying such $P$ and $T$ effects; in contrast to earlier DAC studies[5,6], we did not employ data obtained in a graphite capsule, which causes carbon saturation in silicate melts, since saturation changes the partitioning behaviours[14,15]. Li et al.[16] also did not find the $P$ and $T$ effects on $D_C$ when $\Delta\text{IW}$ < -1.5. In addition, it has been reported[14,16] that the solubilities of carbon in metallic liquid and silicate melt do not depend on pressure. Since the chemical potential of C in a carbon-saturated phase is identical to that in graphite/diamond, it suggests that the pressure dependence of chemical potentials of C is similar among these molten phases and graphite/diamond, and therefore the molar volumes of C in metallic liquid and silicate melt are close to each other, supporting the small pressure dependence of $D_C$ (i.e., small volume change in the reaction).

**Discussion**
**Single-stage core formation model.** The single-stage core formation model is conceptually simple, assuming that the entire core and mantle reached chemical equilibrium under a certain $P$, $T$ and $fO_2$ condition. Such condition has been estimated, based on the metal-silicate partitioning of moderately siderophile elements, to be 40 GPa and 3750 K (ref. [18]), 50 GPa and 3500 K (ref. [19]), or 54 GPa and 3350 K (ref. [20]) (Supplementary Table 1), and $\Delta\text{IW}$ = -2.3. The BSE water abundance may be 710 ppm[23] including 1.6 × 10$^{24}$ g water in the crust, ocean and atmosphere[24], while the higher BSE water contents of 1760 ppm and 1070 ppm $H_2O$ have also been proposed[25,26]. On the other hand, the carbon abundance in the BSE may be 120 ppm[27], 140 ppm[23] or ~160 ppm (or ~220 ppm)[28].

Based on the 710 ppm $H_2O$ (ref. [23]) and 120 ppm C (ref. [27]) in the BSE, we can calculate core concentrations of hydrogen and carbon by using $D_H$ and $D_C$ at the conditions of core-mantle chemical equilibrium (core formation). Following the single-stage core formation model proposed by Wade and Wood[18], Eqs. 3 and 5 showed 0.68 wt% H and 0.75 wt% C in the core with $D_H = 92$ and $D_C = 67$ (Fig. 5a, Supplementary Table 1), in which we employed $nbo/t = 2.75$ that is for a pyrolitic mantle composition[29], fixed the S content to be 2 wt% in the core from cosmochemical and geochemical constraints[27] and calculated the core abundances of Si and O using their metal-silicate partitioning data[18]. Also, similar calculations based on other core formation model[19] including the metal/silicate partition coefficients of Si and O found 0.45 wt% H and 1.15 wt% C in the core ($D_H = 56$ and $D_C = 96$) (Supplementary Table 1). The Fischer et al.[20]'s model provides 0.35 wt% H and 1.29 wt% C with $D_H = 44$ and $D_C = 107$.

These results suggest that the core is a primary reservoir for both hydrogen and carbon; >98% and >97% of their bulk Earth budgets may be present in the core, respectively. We also note that when we employ the separately-determined $D_H$ (ref. [4]) and $D_C$ values[5,6] without considering the strong interaction between hydrogen and carbon, the C content in the core is calculated to be much higher[6] or lower[5] than described above, while H concentration is similar (Fig. 5a). Such core abundances indicate 1.10–2.08 wt% $H_2O$ and 0.26–0.44 wt% C in the bulk Earth (Fig. 5b, Supplementary Table 1), which corresponds to Earth's building blocks if we do not consider volatile loss during accretion.

**Homogeneous accretion & multi-stage core formation model.** The multi-stage core formation models may be more realistic scenarios in which core grows stepwise during the Earth accretion[3]. Here we employ all of the nine multi-stage core formation models reported by Tagawa et al.[4], where the Earth grows by 1000 steps upon accretions of impactors that are identical in size and composition. These models explain ~700 ppm $H_2O$ in the BSE including the ocean water and the mantle FeO, NiO and CoO contents based on their metal/silicate partition coefficients[3,19,20,30] as well as the Earth's core mass fraction. In this study, we additionally examine the distributions of carbon with the newly determined $D_H$ and $D_C$ that consider the effect of their strong interactions (Eqs. 3 and 5).

We explored impactor $H_2O$ and C concentrations, which lead to the amounts of residual water and carbon in silicate to be both within the range of existing estimates of the present-day BSE concentrations of 710–1760 ppm $H_2O$ (refs. [23-26]) and 120–220 ppm

C (refs. [23,27,28]) at the end of accretion. The total difference in mantle concentrations of FeO, NiO and CoO (ref. [31]) and core mass fraction between the modelling results and the target values is defined as:

$$\sigma = \sqrt{1/4 \times \sum \{\frac{(caluculated\ value) - (present\ Earth\ value)}{(error\ of\ the\ present\ Earth\ value)}\}^2} \qquad (6)$$

We suppose that a model explains the core size and the BSE composition reasonably well when $\sigma < 10$. Each model (S1–S3, F1–F3 and R1–R3 models)[4] gives the specific $P$-$T$ evolution of metal-silicate chemical equilibrium, impactor size (= the Earth size divided by $N \times 1000$), impactor bulk (silicate + core) composition (Supplementary Table 2) and exchange coefficients for Ni, Co, O and Si. Here we fixed $\Phi_{Fe}$ (metallic Fe/total Fe) = 0.987 and $\Phi_{Si}$ (metallic Si/total Si) = 0.034, unlike Tagawa et al.[4]'s simulations, by considering the core-mantle partitioning of O and Si in an enstatite-chondrite-like impactor[32] at 1 bar and 2023 K (ref. [29]) using their partition coefficients[30].

We searched for the sets of $P_{final}$ (the pressure of core metal segregation at the final step of accretion) and impactor water and carbon concentrations, which minimize the $\sigma$ value (Eq. 6) while satisfying the residual 710–1760 ppm $H_2O$ and 120–220 ppm C in the BSE. For each of the nine models, we found a $\sigma$ minimum with specific $P_{final}$ (Supplementary Fig. 3). With such $P_{final}$ values, the present homogeneous multi-stage core formation models show that accreting identical impactors with 0.2–2.4 wt% $H_2O$ and 0.2–1.0 wt% C accounts for the BSE composition (FeO, NiO, CoO, $H_2O$ and C contents) and the Earth's core size (Fig. 6a and Supplementary Fig. 4); typical examples of their evolutions during accretion are given in Supplementary Fig. 5. However, the total amount of light impurity elements in the core at the end of the Earth accretion (the calculated core concentrations of H, C, Si and O + assumed 2 wt% S, see Supplementary Fig. 5) are more than required to explain the present-day outer core density deficit[33] (Fig. 6a and Supplementary Fig. 4) even when the temperature at the inner core boundary (ICB) is relatively low ($T_{ICB}$ = 4800 K) and the core liquid is saturated with Si + O under corresponding 3500 K at the core-mantle boundary[34] (note that low $T_{ICB}$ and the $SiO_2$ crystallization from the liquid core leave relatively large room for hydrogen and carbon). It is therefore unlikely from the present modelling that the Earth's building blocks included water and carbon from the beginning of accretion.

Next we examined alternative scenarios in which water and carbon were delivered in a late stage of the Earth accretion. Such a late delivery of volatiles has been considered for sulfur in a previous multi-stage core formation model[35]. In the case of their delivery after 50% accretion (Supplementary Figs. 6–8), the calculated total amount of the core light elements (0.1–0.6 wt% H, 0.1–1.5 wt% C, Si and O contents and assumed S abundance) explains the present-day core density deficit when $T_{ICB}$ is between 5400 K and 4800 K, taking the Si + O saturation and resulting $SiO_2$ crystallization upon core cooling into account[34] (Fig. 6b and Supplementary Fig. 7). These models require >1 wt% $H_2O$ and <1 wt% C in the impactors during the latter half of the accretion in most models. Their typical evolutions are illustrated in Supplementary Fig. 8 for the S1 model showing relatively small σ values when the volatile-rich latter half impactors include 2.0 wt% $H_2O$ and 0.4 wt% C (leading to 1.0 wt% $H_2O$ and 0.2 wt% C in the bulk Earth) (Supplementary Table 3). This specific scenario finds 0.35 wt% H and 0.62 wt% C in the core, indicating that >95% of the bulk Earth budgets of both hydrogen and carbon may be present in the core.

**Heterogeneous accretion & multi-stage core formation model.** It is highly likely that the Earth was formed from a variety of different types of chondritic materials. Indeed, isotopic compositions, in particular Mo (ref. [43]), Ru (ref. [44]) and Zr isotopes[45], suggest that the Earth's building blocks are close to enstatite (E)-chondrite-type materials but involve some carbonaceous chondrites (CCs). The recent isotopic mixing model by Dauphas et al.[46] proposed accretion scenarios to account for the Earth's isotopic composition, showing the proportions and delivery timing of E-, CI- and minor amounts of ordinary (O)- and other carbonaceous (CO+CV) chondrites. Their 4-stage model argued that water- and carbon-rich CC-type materials accreted to our planet mainly in the last 25% accretion, broadly consistent with our modelling described above. Here the Dauphas et al.[46]'s 4-stage accretion scenario was combined with our multi-stage core formation model (model S1[4] in the previous section) such that the model is constrained not only by the BSE elemental composition and the core mass and density deficit but also by the isotopic composition. Furthermore, we also examined the effects of impactor size that changes the efficiency of chemical equilibrium between impactor metal and proto-Earth silicate[4,47] (by varying the number of steps for Earth accretion, $N \times 1000$), differentiation of impactors before accreting to the Earth (by including volatile-free impactors with a frequency of 1-1/n) and volatile loss on Earth.

Considering the wide ranges of 1) the reported $H_2O$ and C contents in E- and CI-chondrite-type materials[2,38-42] (Figs. 7d, e, Supplementary Table 8, Supplementary Fig. 9), 2) $N$ = 1–10 (Supplementary Fig. 10) and 3) n = 1–8 (Supplementary Fig. 11) along with 4) none or 25% carbon loss throughout the Earth accretion (water/hydrogen loss may have been minor)[48,49], we looked for the sets of these parameters that find σ < 10 through the search for $P_{final}$ and 710–1760 ppm $H_2O$ and 120–220 ppm C in the BSE.

Consequently, these modellings show 0.18–0.49 wt% H and 0.19–1.37 wt% C in the core (Fig. 7a). Such core concentrations give 0.53–1.40 wt% $H_2O$ (present as H in the core) and 0.07–0.44 wt% C in the bulk Earth (Fig. 7d). We note that these ranges of the core and bulk Earth H ($H_2O$) and C contents are similar to those obtained by the homogeneous accretion & multi-stage core formation modelling shown above that is based on the model S1 same as here and considers the delivery of water and carbon only after 50% accretion (Fig. 6b). In addition, in almost all cases, the parameter sets that satisfy the BSE composition and the Earth's core size ($\sigma < 10$) provide the total amount of the core light elements accounting for the core density deficit with $T_{ICB}$ = 4800–5400 K.

The 0.19–1.37 wt% C in the core obtained in these simulations is higher than the core C concentrations of ~0.1–0.2 wt% reported by Blanchard et al.[6] and Fischer et al.[5], in which the effect of hydrogen on $D_C$ was not taken into account. It is because the C abundance in the core is almost independent of the $D_C$ in the Blanchard et al.[6]'s model although they used higher $D_C$. The $D_C$ utilized in the modelling by Fischer et al.[5] is the extrapolated one and lower than that applied in this study. Note that the present parameterization of $D_C$ is consistent with the experimental data by Fischer et al.[5] but employs more data including those by Blanchard et al.[6] and this study. In addition, some recent simulations[48,49] included atmosphere as an additional reservoir, which is important in particular for C, but it reduces the calculated core H and C contents only by less than ~5% in the present modelling (Supplementary Fig. 12) since impactor metal reaches chemical equilibrium with a limited portion of an existing magma ocean[47]. Indeed, the Earth accretion and concurrent core formation must have been complicated processes, on which too few constraints are available so far. The present modelling of core composition employed the BSE isotopic composition and the core density deficit as additional constraints, both of which have not been used in earlier simulations. It is noted, however, that the possible ranges of the core H and C abundances demonstrated in this study are still dependent on model assumptions such as the limited efficiency[4] of chemical equilibrium between impactor metal and proto-Earth silicate. This assumes that upon each impact, the timescale of the metal migration to the core is shorter than that of the compositional homogenization of the entire magma ocean[47].

**Origins of Earth's water and carbon.** All of these modellings show that the bulk Earth $H_2O$ (incorporated as H in the core) abundance is higher than that of E-chondrites—in other words, E-chondrites alone cannot be the source of Earth's water (Figs. 5b, 6b and 7d). Furthermore, the heterogeneous multi-stage core formation models show that 1 to 53% of Earth's water may have derived from the non-carbonaceous (NC)-type materials (E- + O-chondrites) and the rest was from the CC-type ones that constitute 7% of the total Earth building blocks[46] (Fig. 7b). Also, 2 to 72% of the Earth's carbon could have originated from the NC-type materials with/without taking 25% loss during accretion into account[48,49] (Fig. 7c), strongly depending on carbon concentration in E-chondrites. We also note that 3–10% and 3–24% of the BSE $H_2O$ and C budgets, respectively, were derived by the late veneer, the additional 0.5% accretion after the core formation[46]. It suggests that majority of Earth's hydrogen and carbon were involved in metal-silicate partitioning, which causes isotopic fractionation leaving their isotopic ratios in the BSE different from those of the bulk Earth and Earth building blocks.

**Methods**

**High-pressure melting experiments.** Laser-heated DAC techniques were used to generate high *P-T* conditions (Table 1). We prepared five different glasses for silicate starting materials, which have mid-ocean ridge basalt (MORB) compositions with various amounts of $H_2O$; 1.0 wt% (w01), 0.15 wt% (w02), 0.6 wt% (w03), and none (n01 and n02) (Supplementary Table 4). A thin foil (~7 µm thick) of iron containing 4.0 wt% carbon, same as that employed previously[50], was sandwiched between layers of silicate glass. They were loaded into a 120 µm hole at the centre of a pre-indented rhenium gasket. Diamond anvils with flat 300 µm culet were used. After loading, for experiments with anhydrous glass starting materials, we put the DAC into a vacuum oven at 400 K for >12 hours to remove moisture on the sample and subsequently compressed the sample in an argon atmosphere to avoid water adsorption. Under high pressure of interest, the sample was heated from both sides with a couple of 100 W single-mode Yb fibre lasers at BL10XU, SPring-8 synchrotron radiation source with *in-situ* XRD measurements[51]. XRD patterns were collected before/during/after heating using a monochromatic X-ray beam with an energy of ~30 keV that was focused to 6 µm area (full width of half maximum) on the sample position. Melting of a sample was confirmed by the disappearance of XRD peaks except weak ones that derived from silicate liquidus phases outside of a silicate melt pool (Fig. 2). Heating duration ranged from 4 to 60 seconds, which is long enough for each element to diffuse in liquid metal[52] and silicate melt[53]. Indeed, both metal and silicate were found to be homogeneous in composition in this study, except for heterogeneous distributions of small metallic blobs in the silicate melt pool (Fig. 1b). It has been demonstrated by earlier experiments using a multi-anvil press that molten metal and silicate melt reached chemical equilibrium in such heating time scale[54]. Considering much smaller sample size and higher temperatures in the present DAC study than in the multi-anvil experiments, the time necessary for chemical equilibration should be shorter. Temperature was measured by a spectro-radiometric method[51]. Pressure was measured based on the Raman shift of a diamond anvil at 300 K after heating[55] and corrected for thermal pressure contributions that have been estimated to be +2.5 GPa per 1000 K (refs. [4,56]). The errors in pressure and temperature may be ±10% and ±5%, respectively[57].

**XRD and EPMA analyses of liquid metal.** The hydrogen contents in liquid iron were estimated from XRD data collected at high pressures and 300 K since iron sample loses hydrogen when it transforms into the body-centred cubic structure[4,58]. The XRD patterns showed the appearance of face-centred-cubic (fcc) FeH$_x$ and hydrogen-bearing (FeH$_x$)$_7$C$_3$ [or (FeH$_x$)$_3$C] formed from liquid metal upon quenching temperature (Fig. 2c). Their lattice volumes were larger than those of pure fcc Fe and Fe$_7$C$_3$ (or Fe$_3$C) at equivalent pressure and 300 K, respectively[59-61], which is attributed to the incorporation of hydrogen into their interstitial sites. Following Tagawa et al.[4], the hydrogen content $x$ was calculated as:

$$x = \frac{V_{\text{FeH}_x} - V_{\text{Fe}}}{\Delta V_{\text{H\_Fe}}} \text{ in FeH}_x \tag{7}$$

$$x = \frac{V_{(\text{FeH}_x)_7\text{C}_3} - V_{\text{Fe}_7\text{C}_3}}{\Delta V_{\text{H\_Fe7C3}}} \text{ in (FeH}_x)_7\text{C}_3 \tag{8}$$

$$x = \frac{V_{(\text{FeH}_x)_3\text{C}} - V_{\text{Fe}_3\text{C}}}{\Delta V_{\text{H\_Fe3C}}} \text{ in (FeH}_x)_3\text{C} \tag{9}$$

where $V_{\text{FeH}x}$, $V_{(\text{FeH}x)7\text{C}3}$ and $V_{(\text{FeH}x)3\text{C}}$ represent the observed lattice volume per Fe atom of each phase, and $V_{\text{Fe}}$, $V_{\text{Fe7C3}}$ and $V_{\text{Fe3C}}$ are from their equations of state[59-61]. $\Delta V_{\text{H}}$ represents the volume increase per hydrogen atom. We employed $\Delta V_{\text{H\_Fe}}$ previously reported for fcc Fe[62]. $\Delta V_{\text{H\_Fe7C3}}$ was determined by first-principles calculations in this study (see below for the first-principles calculations section, Supplementary Table 5 and Supplementary Fig. 13). Since hydrogen favours different interstitial sites depending on hydrogen concentration, we calculated $x$ in (FeH$_x$)$_7$C$_3$ based on three different $\Delta V_{\text{H\_Fe7C3}}$ applicable to each hydrogen concentration range, $x < 0.14$, $0.14 < x < 0.29$ and $0.29 < x < 0.57$ (Supplementary Table 6). We also found that $\Delta V_{\text{H}}$ is similar between Fe and Fe$_7$C$_3$, in particular $x = 0.22$ and 0.45 observed in (FeH$_x$)$_3$C at 45–47 GPa and 300 K (Supplementary Table 6 and Supplementary Fig. 13) and therefore used $\Delta V_{\text{H\_Fe7C3}}$ to calculate $x$ in (FeH$_x$)$_3$C. In all runs, fcc FeH$_x$, ε-FeOOH and (FeH$_x$)$_7$C$_3$ [or (FeH$_x$)$_3$C] were found in XRD patterns collected from temperature-quenched liquid. Their hydrogen concentrations $x$ were estimated to be 0.03–1.76, 0.21–0.30, and 0.22–0.45, respectively (Supplementary Table 6). We then calculated the total hydrogen concentration in liquid to be 0.33 to 0.98 wt% (Supplementary Table 7) based on the proportions of these crystals in quenched liquid, which were obtained by mass balance calculations using the FE-EPMA analyses (see below) considering the presence of silicon in the iron site. The incorporation of silicon into Fe increases its unit-cell volume to a minor extent[63]. While 0.16–6.47 wt% Si may be present in these iron and iron carbides in the present experiments (Supplementary Table 7), the estimate of the hydrogen content $x$ is reduced by 0.02 at 50 GPa even when 5.9 wt% Si is included[63].

The carbon contents in quenched metal liquids were obtained based on the electron microprobe analyses on the cross sections of recovered samples. After recovering samples at ambient conditions, sample cross sections at the centre of a laser-heated spot were prepared parallel to the compression/laser-heating axis with a focused ion beam (FIB, Versa 3D™, FEI). Textural and chemical characterizations were made with an FE-type scanning electron microscope (FE-SEM) and energy dispersive X-ray spectroscopy (EDS). Gold coating was made. We found 3.04–9.51 wt% C in quenched liquid metal along with other major elements with an FE-EPMA (JXA-8530F, JEOL) (Supplementary Table 7). At least five different points were measured for each phase. We employed acceleration voltage of 12 kV, beam current of 15 nA, and LIF (Fe), TAP (Al, Na), TAPH (Mg), LDE1 (O), LDE2H (C) and PETJ (K, Ca, Si, Ti) as analyzing crystals. The calibration curve for C was obtained with $Fe_3C$, Fe-0.84 wt% C (JSS066-6, the Japan Iron and Steel Federation), and rhenium gasket (assumed to be free of carbon) (Supplementary Fig. 14). Fe, $Al_2O_3$, MgO, Si, $SiO_2$, $TiO_2$, $CaSiO_3$, $KTiOPO_4$, and $NaAlSi_3O_8$ were used as standards for other elements. The X-ray counting time for peak/background was 20/10 seconds. The ZAF correction was applied. The major element compositions of silicate melt and the silicate starting materials were also obtained with an FE-EPMA (Supplementary Table 4).

**SIMS measurements of silicate melt.** We determined the hydrogen (water) and carbon contents in quenched silicate melts with SIMS equipped with a two-dimensional ion detector, stacked CMOS-type active pixel sensor (SCAPS) at the Hokkaido University[4,64-67] (Figs. 1c, d). This system provides quantitative maps of secondary ions emitted from the sample surface because the CMOS sensor exhibits a good linear relationship between an output voltage and the number of secondary ions[66]. Therefore, the abundance of each element from the intensity map can be quantified. The sample cross sections were coated with a ~70 nm thick Au layer by plasma sputtering to compensate for electrostatic charging on the sample surface. We used the $Cs^+$ primary beam rastered across a 100 μm × 100 μm region on the sample. In order to prevent the effect of electrical charging, analyses were performed while applying electrons with an electron gun. A contrast aperture was set to be 100 μm in diameterand the exit slit was opened fully for 1H⁻ and 750 μm for others. Pressure during measurement was $2 \times 10^{-9}$ Pa. The SCAPS images of $^1H^-$, $^{12}C^-$ and $^{28}Si^-$ secondary ions were collected. Accumulation time was 150, 100 and 10 seconds in each image, respectively. We employed five silicate glasses with known $H_2O$ and C concentrations as standards (0.0 to 1.8 wt%, 0.0 to 389 ppm, respectively)[68] and prepared an additional silicate glass standard that contains 1.3 wt% C to fully cover the range of C concentration in silicate melt in the present experiments. Based on these standard glass analyses, we converted intensity ratios, $^1H^-/^{28}Si^-$ and $^{12}C^-/^{28}Si^-$ into mass ratios. The intensity ratios from the secondary ion images are linearly correlated with the mass ratios of H/Si or C/Si of these standards (correlation coefficient $R^2$ = 0.997 and 1.000, respectively, as shown in Supplementary Fig. 15). Water (hydrogen) or carbon concentration in silicate melt was then obtained by multiplying the H/Si or C/Si ratio by the Si content acquired by FE-EPMA. While silicate melt in run #1 included $H_2O$ much more than 1.8 wt% in the standard glass, each parameter in Eq. 3 for $D_H$ does not change beyond uncertainty when excluding the data from run #1.

Hydrogen distributions were homogeneous in quenched silicate melt (Supplementary Fig. 16), suggesting that hydrogen did not migrate from liquid metal to surrounding silicate upon quenching temperature. The $^{12}C^-$ intensities from metal were higher by more than two orders of magnitude than those from silicate melt (Supplementary Fig. 16). The high C intensity from the metal leaked into the surrounding silicate melt due to the lens-flare or aberration effect. Therefore, the regions of interest (ROIs) in the silicate melts, from which we obtained their hydrogen and carbon abundances, were chosen from an area free from such flare effect from the metal based on the 2D map and line profile of the $\log_{10}(C)$ values (Supplementary Fig. 16). In addition, we avoided an area where small metallic blobs are present in the silicate melt. As shown in a back-scattered electron image (Fig. 1b), the outer part of a silicate melt layer is free from the metallic blobs. Following the arguments made in the earlier experiments on metal-silicate partitioning of carbon by Blanchard et al.[6], we interpret that such blobs are not quench products but were present during melting. Indeed, the back-scattered electron image given in Fig. 1b does not show the Fe depletion in areas next to the blobs, which supports that they were not formed from silicate melt upon quenching. The SIMS analyses demonstrate that the amounts of $H_2O$ widely ranged from 0.11 to 10.78 wt% in silicate melts (Supplementary Table 4). Their carbon contents were found to be 0.02–0.60 wt% and 0.008–0.44 wt% when using hydrous and anhydrous silicate starting materials, respectively.

**Activities of elements in (H, C)-rich metal and oxygen fugacity.** Liquid metals found in this study included carbon (and hydrogen in experiments using $H_2O$-bearing starting materials). The presence of hydrogen and carbon decreases the molar fraction of iron in metal ($x_{Fe}^{metal}$), which apparently increases the oxygen fugacity relative to iron-wüstite (IW) buffer, $\Delta IW \sim 2 \, \log_{10}(x_{FeO}^{silicate}/x_{Fe}^{metal})$ and changes the exchange coefficient $K_D^O = x_{Fe}^{metal} x_O^{metal}/x_{FeO}^{silicate}$ for the reaction $FeO^{silicate} = Fe^{metal} + O^{metal}$. According to Tagawa et al.[4], the $K_D^O$ values calculated without considering the presence of hydrogen and carbon are consistent with those obtained in the (H, C)-free system. It suggests that both hydrogen and carbon do not have colligative properties in iron solvent since small H and C atoms are incorporated into liquid Fe interstitially. Therefore, the activity of element $i$ in metal is approximated as;

$$x_i' = \frac{N_i}{\sum_{k \neq H,C} N_k} . \quad (10)$$

We obtained the mol-based $D$ values from earlier experimental data with the same procedure.

**First-principles calculations.** First-principles calculations based on density functional theory were performed using the Quantum Espresso (QE) codes[69]. Generalized-gradient approximation[70] was adopted, along with projector-augmented wave (PAW) pseudopotentials available on the QE website (energy cutoff of 90 Ry). In our calculations, $Fe_7C_3$ (20-atom cells, $P6_3mc$ symmetry) with various site occupations of H were considered, including Wyckoff positions $2a$, $2b$, $6c$ and $12d$. All considered structures are fully optimized (4×4×6 k-point mesh), and the results are fitted to the 3rd-order Birch–Murnaghan equation of state (Supplementary Table 5, Supplementary Fig. 13). For all considered structures, ferromagnetic (FM) state is more favorable than the nonmagnetic (NM) state. As the H concentration increases, the favorable site occupations are $2a$, $2a+2b$ and $2a+6c$ for $Fe_7C_3H$, $Fe_7C_3H_2$ and $Fe_7C_3H_4$, respectively.

**References**

1. Alexander, C. M. O. et al. The provenances of asteroids, and their contributions to the volatile inventories of the terrestrial planets. *Science* **337**, 721–723 (2012).
2. Piani, L. et al. Earth's water may have been inherited from material similar to enstatite chondrite meteorites. *Science* **369**, 1110–1113 (2020).

3. Rubie, D. C. et al. Accretion and differentiation of the terrestrial planets with implications for the compositions of early-formed Solar System bodies and accretion of water. *Icarus* **248**, 89–108 (2015)
4. Tagawa, S. et al. Experimental evidence for hydrogen incorporation into Earth's core. *Nat. Commun.* **12**, 2588 (2021).
5. Fischer, R. A., Cottrell, E., Hauri, E., Lee, K. K. M. & Le Voyer, M. The carbon content of Earth and its core. *Proc. Natl. Acad. Sci. USA* **117**, 8743–8749 (2020).
6. Blanchard, I. et al. The metal–silicate partitioning of carbon during Earth's accretion and its distribution in the early solar system. *Earth Planet. Sci. Lett.* **580**, 117374 (2022).
7. Kuwayama, Y. et al. Equation of state of liquid iron under extreme conditions. *Phys. Rev. Lett.* **124**, 165701 (2020).
8. Hirose, K., Wood, B. & Vočadlo, L. Light elements in the Earth's core. *Nat. Rev. Earth Environ.* **2**, 645–658 (2021).
9. Marty, B. The origins and concentrations of water, carbon, nitrogen and noble gases on Earth. *Earth Planet. Sci. Lett.* **313-314**, 56–66 (2012).
10. Hirschmann, M. M. Constraints on the early delivery and fractionation of Earth's major volatiles from C/H, C/N, and C/S ratios. *Am. Mineral.* **101**, 540–553 (2016).
11. Li, Y., Vočadlo, L., Sun, T. & Brodholt, J. P. The Earth's core as a reservoir of water. *Nat. Geosci.* **13**, 453–458 (2020).
12. Yuan, L. & Steinle-Neumann, G. Strong sequestration of hydrogen into the Earth's core during planetary differentiation. *Geophys. Res. Lett.* **47**, e2020GL088303 (2020).
13. Dasgupta, R., Chi, H., Shimizu, N., Buono, A. S. & Walker, D. Carbon solution and partitioning between metallic and silicate melts in a shallow magma ocean: implications for the origin and distribution of terrestrial carbon. *Geochim. Cosmochim. Acta* **102**, 191–212 (2013).
14. Kuwahara, H., Itoh, S., Suzumura, A., Nakada, R. & Irifune, T. Nearly carbon-saturated magma oceans in planetary embryos during core formation. *Geophys. Res. Lett*. **48**, e2021GL092389 (2021).
15. Kuwahara, H., Itoh, S., Nakada, R. & Irifune, T. The effects of carbon concentration and silicate composition on the metal-silicate partitioning of carbon in a shallow magma ocean. *Geophys. Res. Lett.* **46**, 9422–9429 (2019).
16. Li, Y., Dasgupta, R., Tsuno, K., Monteleone, B. & Shimizu, N. Carbon and sulfur budget of the silicate Earth explained by accretion of differentiated planetary embryos. *Nat. Geosci.* **9**, 781–785 (2016).

17. Fichtner, C. E., Schmidt, M. W., Liebske, C., Bouvier, A.-S. & Baumgartner, L. P. Carbon partitioning between metal and silicate melts during Earth accretion. *Earth Planet. Sci. Lett.* **554**, 116659 (2021).
18. Wade, J. & Wood, B. J. Core formation and the oxidation state of the Earth. *Earth Planet. Sci. Lett.* **236**, 78–95 (2005).
19. Siebert, J., Badro, J., Antonangeli, D. & Ryerson, F. J. Metal–silicate partitioning of Ni and Co in a deep magma ocean. *Earth Planet. Sci. Lett.* **321-322**, 189–197 (2012).
20. Fischer, R. A. et al. High pressure metal–silicate partitioning of Ni, Co, V, Cr, Si, and O. *Geochim. Cosmochim. Acta* **167**, 177–194 (2015).
21. Ma, Z. Thermodynamic description for concentrated metallic solutions using interaction parameters. *Metall. Mater. Trans. B* **32**, 87–103 (2001).
22. Siebert, J., Corgne, A. & Ryerson, F. J. Systematics of metal–silicate partitioning for many siderophile elements applied to Earth's core formation. *Geochim. Cosmochim. Acta* **75**, 1451–1489 (2011).
23. Hirschmann, M. M. Comparative deep Earth volatile cycles: the case for C recycling from exosphere/mantle fractionation of major ($H_2O$, C, N) volatiles and from $H_2O$/Ce, $CO_2$/Ba, and $CO_2$/Nb exosphere ratios. *Earth Planet. Sci. Lett.* **502**, 262–273 (2018).
24. Lécuyer, C., Gillet, P. & Robert, F. The hydrogen isotope composition of seawater and the global water cycle. *Chem. Geol.* **145**, 249–261 (1998).
25. Peslier, A. H., Schönbächler, M., Busemann, H. & Karato, S.-I. Water in the Earth's interior: distribution and origin. *Space Sci. Rev.* **212**, 1–68 (2017).
26. Palme, H. & O'Neill, H. S. C. Cosmochemical estimates of mantle composition. In *Treatise on Geochemistry* 2nd edn, Vol. 3 (eds. Holland, H. D. & Turekian, K. K.) 1–39 (Elsevier, 2014).
27. McDonough, W. F. Compositional model for the Earth's core. In *Treatise on Geochemistry* 2nd edn, Vol. 3 (eds. Holland, H. D. & Turekian, K. K.) 559–577 (Elsevier, 2014).
28. Marty, B. et al. An evaluation of the C/N ratio of the mantle from natural $CO_2$-rich gas analysis: geochemical and cosmochemical implications. *Earth Planet. Sci. Lett.* **551**, 116574 (2020).
29. Takahashi, E. Melting of a dry peridotite KLB-1 up to 14 GPa: implications on the origin of peridotitic upper mantle. *J. Geophys. Res.* **91**, 9367–9382 (1986).
30. Siebert, J., Badro, J., Antonangeli, D. & Ryerson, F. J. Terrestrial accretion under oxidizing conditions. *Science* **339**, 1194–1197 (2013).

31. Wang, H. S., Lineweaver, C. H. & Ireland, T. R. The elemental abundances (with uncertainties) of the most Earth-like planet. *Icarus* **299**, 460–474 (2018).
32. Javoy, M. et al. The chemical composition of the Earth: enstatite chondrite models. *Earth Planet. Sci. Lett.* **293**, 259–268 (2010).
33. Umemoto, K. & Hirose, K. Chemical compositions of the outer core examined by first principles calculations. *Earth Planet. Sci. Lett.* **531**, 116009 (2020).
34. Hirose, K. et al. Crystallization of silicon dioxide and compositional evolution of the Earth's core. *Nature* **543**, 99–102 (2017).
35. Suer, T.-A., Siebert, J., Remusat, L., Menguy, N. & Fiquet, G. A sulfur-poor terrestrial core inferred from metal–silicate partitioning experiments. *Earth Planet. Sci. Lett.* **469**, 84–97 (2017).
36. Hirschmann, M. M., Bergin, E. A., Blake, G. A., Ciesla, F. J. & Li, J. Early volatile depletion on planetesimals inferred from C–S systematics of iron meteorite parent bodies. *Proc. Natl Acad. Sci. USA* **118**, e2026779118 (2021).
37. Newcombe, M. E. et al. Degassing of early-formed planetesimals restricted water delivery to Earth. *Nature* **615**, 854–857 (2023).
38. Lodders, K. Relative atomic solar system abundances, mass fractions, and atomic masses of the elements and their isotopes, composition of the solar photosphere, and compositions of the major chondritic meteorite groups. *Space Sci. Rev.* **217**, 44 (2021).
39. Wasson, J. T. & Kallemeyn, G. W. Compositions of chondrites. *Philos. T. R. Soc. A* **325**, 535–544 (1988).
40. Moore, C. B. & Lewis, C. F. The distribution of total carbon content in enstatite chondrites. *Earth Planet. Sci. Lett.* **1**, 376–378 (1966).
41. Alexander, C. M. O. Quantitative models for the elemental and isotopic fractionations in the chondrites: the non-carbonaceous chondrites. *Geochim. Cosmochim. Acta* **254**, 246–276 (2019).
42. Alexander, C. M. O. Quantitative models for the elemental and isotopic fractionations in chondrites: the carbonaceous chondrites. *Geochim. Cosmochim. Acta* **254**, 277–309 (2019).
43. Budde, G., Burkhardt, C. & Kleine, T. Molybdenum isotopic evidence for the late accretion of outer Solar System material to Earth. *Nat. Astron.* **3**, 736–741 (2019).
44. Fischer-Gödde, M. et al. Ruthenium isotope vestige of Earth's pre-late-veneer mantle preserved in Archaean rocks. *Nature* **579**, 240–244 (2020).
45. Burkhardt, C. et al. Terrestrial planet formation from lost inner Solar System material. *Sci. Adv.* **7**, eabj7601 (2021).

46. Dauphas, N., Hopp, T. & Nesvorný, D. Bayesian inference on the isotopic building blocks of Mars and Earth. *Icarus* **408**, 115805 (2024).
47. Deguen, R., Olson, P. & Cardin, P. Experiments on turbulent metal–silicate mixing in a magma ocean. *Earth Planet. Sci. Lett.* **310**, 303–313 (2011).
48. Sakuraba, H., Kurokawa, H., Genda, H. & Ohta, K. Numerous chondritic impactors and oxidized magma ocean set Earth's volatile depletion. *Sci. Rep.* **11**, 20894 (2021).
49. Gu, J. T. et al. Composition of Earth's initial atmosphere and fate of accreted volatiles set by core formation and magma ocean redox evolution. *Earth Planet. Sci. Lett.* **629**, 118618 (2024).
50. Mashino, I., Miozzi, F., Hirose, K., Morard, G. & Sinmyo, R. Melting experiments on the Fe–C binary system up to 255 GPa: constraints on the carbon content in the Earth's core. *Earth Planet. Sci. Lett.* **515**, 135–144 (2019).
51. Hirao, N. et al. New developments in high-pressure X-ray diffraction beamline for diamond anvil cell at SPring-8. *Matter Radiat. Extrem.* **5**, 018403 (2020).
52. Helffrich, G. Outer core compositional layering and constraints on core liquid transport properties. *Earth Planet. Sci. Lett.* **391**, 256–262 (2014).
53. de Koker, N. P., Stixrude, L. & Karki, B. B. Thermodynamics, structure, dynamics, and freezing of $Mg_2SiO_4$ liquid at high pressure. *Geochim. Cosmochim. Acta* **72**, 1427–1441 (2008).
54. Thibault, Y. & Walter, M. J. The influence of pressure and temperature on the metal-silicate partition coefficients of nickel and cobalt in a model C1 chondrite and implications for metal segregation in a deep magma ocean. *Geochim. Cosmochim. Acta* **59**, 991–1002 (1995).
55. Akahama, Y. & Kawamura, H. High-pressure Raman spectroscopy of diamond anvils to 250 GPa: method for pressure determination in the multimegabar pressure range. *J. Appl. Phys.* **96**, 3748–3751 (2004).
56. Andrault, D. et al. Solidus and liquidus profiles of chondritic mantle: implication for melting of the Earth across its history. *Earth Planet. Sci. Lett.* **304**, 251–259 (2011).
57. Mori, Y. et al. Melting experiments on Fe–$Fe_3S$ system to 254 GPa. *Earth Planet. Sci. Lett.* **464**, 135–141 (2017).
58. Fukai, Y. & Suzuki, T. Iron-water reaction under high pressure and its implication in the evolution of the Earth. *J. Geophys. Res.* **91**, 9222–9230 (1986).
59. Dorogokupets, P. I., Dymshits, A. M., Litasov, K. D. & Sokolova, T. S. Thermodynamics and equations of state of iron to 350 GPa and 6000 K. *Sci. Rep.* **7**,

41863 (2017).

60. Nakajima, Y. et al. Thermoelastic property and high-pressure stability of $Fe_7C_3$: implication for iron-carbide in the Earth's core. *Am. Mineral.* **96**, 1158–1165 (2011).

61. McGuire, C. et al. *P-V-T* measurements of $Fe_3C$ to 117 GPa and 2100 K: implications for stability of $Fe_3C$ phase at core conditions. *Am. Mineral.* **106**, 1349–1359 (2021).

62. Tagawa, S., Gomi, H., Hirose, K. & Ohishi, Y. High-temperature equation of state of FeH: implications for hydrogen in Earth's inner core. *Geophys. Res. Lett.* **49**, e2021GL096260 (2022).

63. Fischer, R. A. et al. Equations of state in the Fe-FeSi system at high pressures and temperatures. *J. Geophys. Res.* **119**, 2810–2827 (2014).

64. Tsutsumi, Y. et al. Retention of water in subducted slabs under core–mantle boundary conditions. *Nat. Geosci.* **17**, 697–704 (2024).

65. Yurimoto, H., Nagashima, K. & Kunihiro, T. High precision isotope micro-imaging of materials. *Appl. Surf. Sci.* **203–204**, 793–797 (2003).

66. Sakamoto, N. et al. Remnants of the early solar system water enriched in heavy oxygen isotopes. *Science* **317**, 231–233 (2007).

67. Greenwood, J. P. et al. Hydrogen isotope ratios in lunar rocks indicate delivery of cometary water to the Moon. *Nat. Geosci.* **4**, 79–82 (2011).

68. Shimizu, K. et al. $H_2O$, $CO_2$, F, S, Cl, and $P_2O_5$ analyses of silicate glasses using SIMS: report of volatile standard glasses. *Geochem. J.* **51**, 299–313 (2017).

69. Giannozzi, P. et al. Advanced capabilities for materials modelling with Quantum ESPRESSO. *J. Phys. Condens. Matter* **29**, 465901 (2017).

70. Perdew, J.P., Burke, K. & Ernzerhof, M. Generalized gradient approximation made simple. *Phys. Rev. Lett.* **78**, 1396 (1997).

71. Yokoyama, T. et al. Samples returned from the asteroid Ryugu are similar to Ivuna-type carbonaceous meteorites. *Science* **379**, eabn7850 (2023).

72. Lauretta, D. et al. Asteroid (101955) Bennu in the laboratory: properties of the sample collected by OSIRIS-REx. *Meteorit. Planet. Sci.* **59,** 2453–2486 (2024).

**Acknowledgments** The authors acknowledge K. Yonemitsu for assisting EPMA analyses. We thank S. Kawaguchi and H. Kadobayashi for their help in experiments at BL10XU, SPring-8 (proposal no. 2021B0181). We also thank S. Tagawa for his advice on melting experiments. Comments from four anonymous referees greatly helped improve the manuscript. This work was supported by the JSPS Kakenhi 21H04968 to K.H., 23KJ0479 to Y.T. and 21H04985 to H.Y. and by the "Imaging Platform" program by MEXT. H.H. acknowledges supports from the National Science and Technology Council of Taiwan under Grant No. NSTC 113-2116-M-008-010.

**Author contributions** Y.T. and K.H. designed and led the project. This study is based on DAC experiments by Y.T. and S.Y., SIMS analyses by N.S., H.Y. and Y.T., first-principles calculations by H.H., and computational modelling by Y.T., S.M. and S.Y. Y.T. and K.H. wrote the manuscript, and all authors commented on it.

**Competing interests** The authors declare no competing interests.
**Correspondence and requests for materials** should be addressed to K.H.

**Table 1 Experimental conditions and results.**

| Run # | Pressure (GPa) | Temperature (K) | Duration (second) | Starting materials | $fO_2$ ($\Delta$IW) | nbo/t | $D_H^{metal/silicate}$ | $D_C^{metal/silicate}$ |
|---|---|---|---|---|---|---|---|---|
| 1 | 42 (4) | 3940 (200) | 10 | w01+Fe-4wt%C | -1.74 | 3.51 | 0.7 (2) | 27 (8) |
| 2 | 55 (6) | 4760 (240) | 10 | w03+Fe-4wt%C | -1.31 | 1.69 | 26.1 (63) | 132 (53) |
| 3 | 56 (6) | 3800 (190) | 9 | w02+Fe-4wt%C | -1.36 | 1.74 | 8.2 (11) | 49 (12) |
| 4 | 56 (6) | 4130 (210) | 10 | w01+Fe-4wt%C | -1.76 | 0.82 | 6.3 (19) | 36 (23) |
| 5 | 55 (5) | 3630 (180) | 30 | w01+Fe-4wt%C | -1.78 | 0.90 | 3.1 (8) | 41 (8) |
| 6 | 33 (3) | 4060 (200) | 10 | n02+Fe-4wt%C | -1.25 | 1.46 | | 84 (25) |
| 7 | 43 (4) | 3930 (200) | 10 | n01+Fe-4wt%C | -0.93 | 1.77 | | 77 (35) |
| 8 | 41 (4) | 3660 (180) | 10 | n02+Fe-4wt%C | -1.26 | 1.01 | | 141 (62) |
| 9 | 46 (5) | 3980 (200) | 60 | n02+Fe-4wt%C | -2.09 | 0.51 | | 369 (324) |
| 10 | 54 (5) | 4060 (200) | 4 | n01+Fe-4wt%C | -0.60 | 2.24 | | 14 (4) |

$D_H$ and $D_C$ are based on molar concentrations.

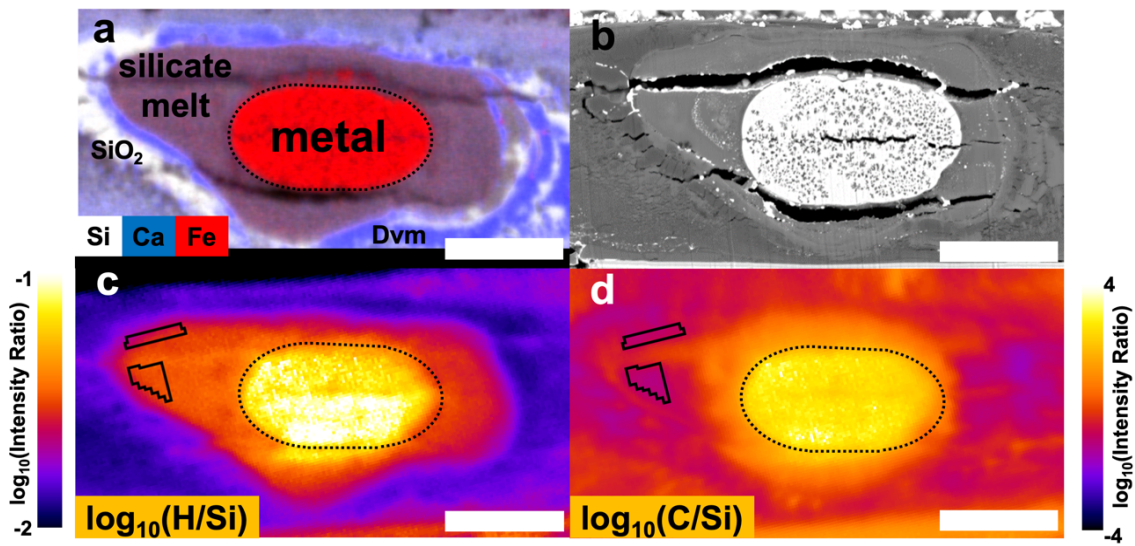

**Fig. 1 Sample cross section from run #1. a** Combined X-ray maps of Si (white), Ca (blue) and Fe (red), showing quenched liquid metal, silicate melt, $CaSiO_3$ davemaoite (Dvm) and the $SiO_2$ phase. **b** Back-scattered electron (BSE) image. **c, d** The SCAPS images of secondary ions intensity ratios of H/Si and C/Si. The H and C contents in silicate melt were determined from the regions of interest (ROIs, enclosed by black lines in **c** and **d**) in an area free from the lens-flare effects of the secondary ion optics (see Supplementary Fig. 16 and "Methods") and metal blobs (see the BSE image in **b**). A quenched metal portion in **c** and **d** is from the X-ray map in **a**. Note that **a**, **c/d**, and **b** are slightly different cross sections because of sputtering by the SIMS ion beam and repolishing with a focused ion beam in between. Scale bar, 10 μm.

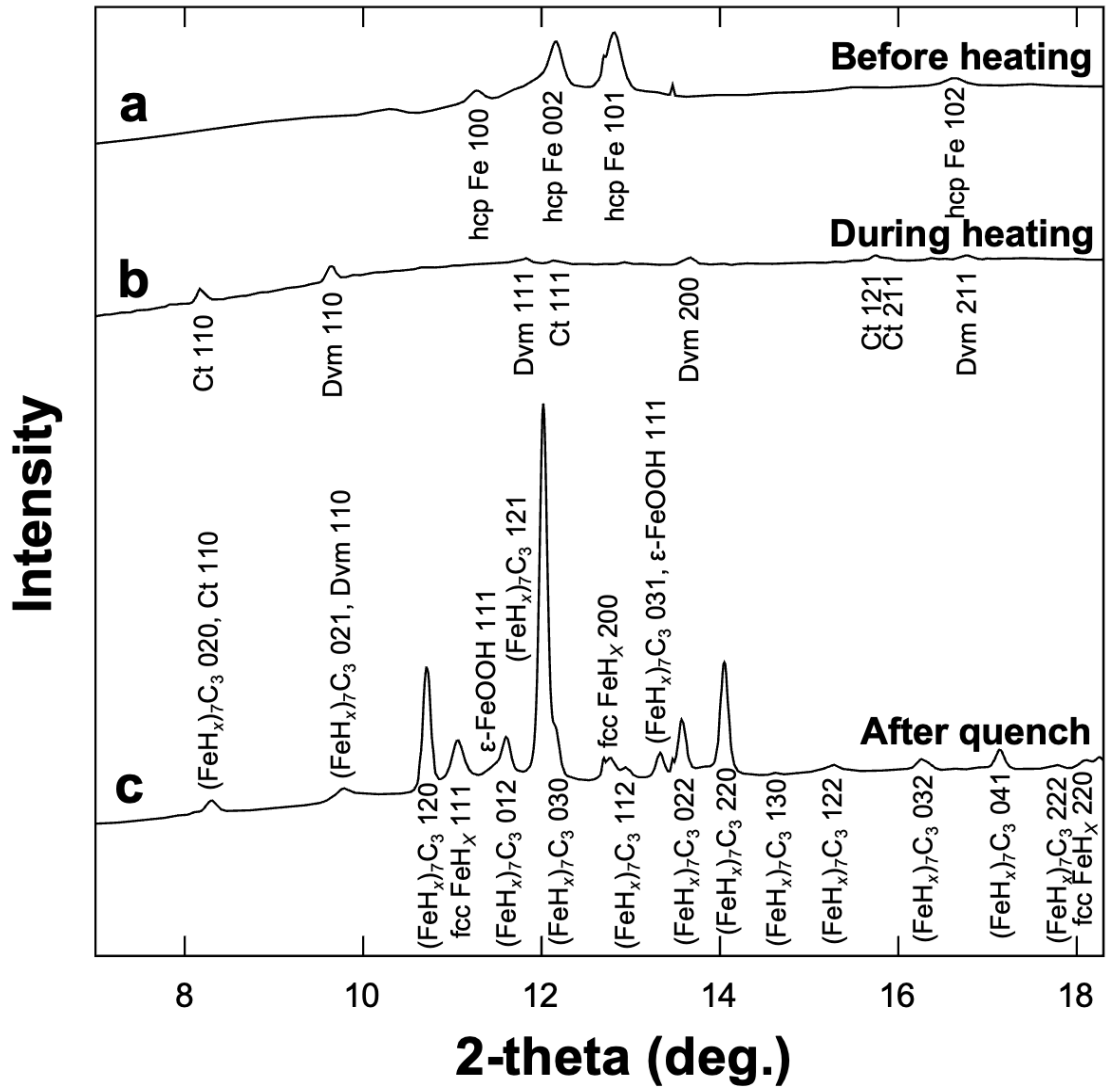

**Fig. 2 XRD patterns obtained in high-pressure melting experiment.** They were collected before (**a**), during (**b**) and after heating (**c**) to 3940 K at 42 GPa in run #1. During heating, XRD peaks were observed only from Dvm and CaCl$_2$-type SiO$_2$ (Ct), which were present outside of a silicate melt pool (see Fig. 1a), indicating the melting of both metal and silicate. After quenching temperature, the peaks from fcc FeH$_x$, ε-FeOOH and (FeH$_x$)$_7$C$_3$ appeared. The amount of hydrogen in liquid metal was estimated from the phase proportion and hydrogen concentrations in these quench crystals. See text for details.

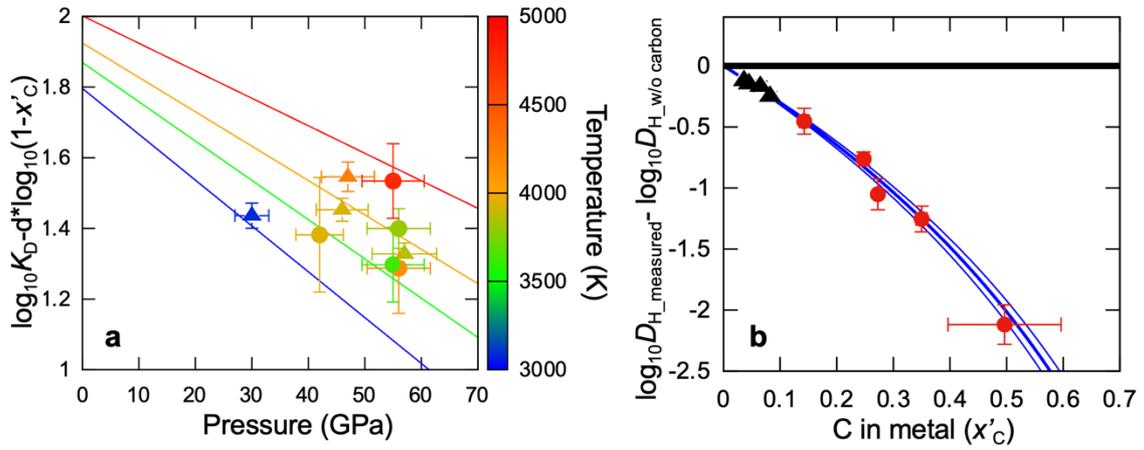

**Fig. 3 Partitioning of hydrogen. a** Pressure and temperature dependence of the exchange coefficient $K_D$ for metal-silicate partitioning of hydrogen after correcting for the effect of carbon (Eq. 3). Blue (3000 K), light green (3500 K), orange (4000 K) and red lines (5000 K) show isotherms. **b** Reduction in metal/silicate partition coefficient $D$ of hydrogen (molar basis) as a function of carbon concentration in liquid metal. Circles, this study; triangles, Tagawa et al.[4]

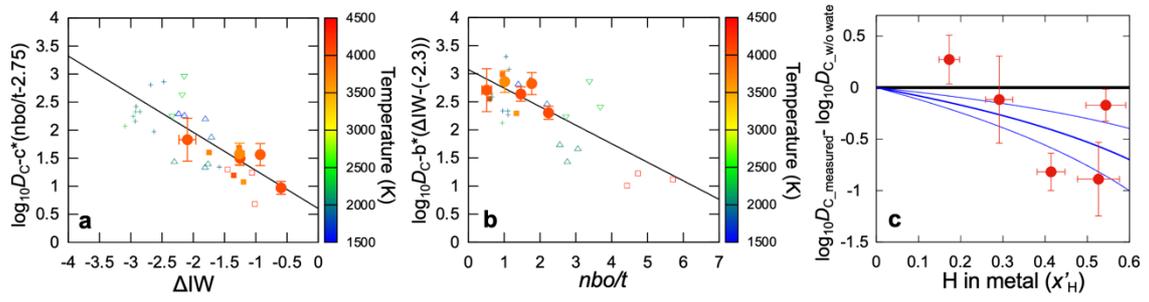

**Fig. 4 Partitioning of carbon.** Partition coefficient $D$ of carbon (molar basis) as a function of $\Delta$IW (**a**) and $nbo/t$ (**b**) obtained in hydrogen-free experiments and of hydrogen concentration in metal (**c**). See Eq. 5. Color indicates temperature. Circles, this study; squares, Blanchard et al.[6] (closed) and Fischer et al.[5] (open); triangles, Fichtner et al.[17] (normal) and Dasgupta et al.[13] (inverted); crosses, Kuwahara et al.[14,15]

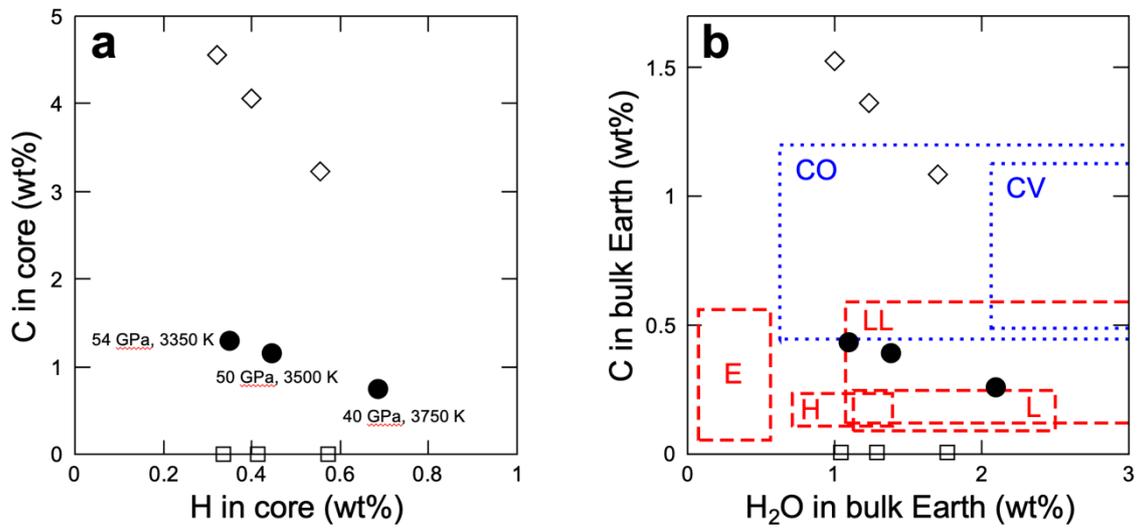

**Fig. 5 Modelling of single-stage core formation.** Hydrogen (water) and carbon concentrations in the core (**a**) and the bulk Earth (**b**) considering the *P-T* conditions of core metal segregation at 40 GPa and 3750 K[18], 50 GPa and 3500 K[19], and 54 GPa and 3350 K[20]. Black circles, calculated using $D_H$ and $D_C$ simultaneously determined in this study; diamonds, $D_H$ from Tagawa et al.[4] and $D_C$ from Blanchard et al.[6]; squares, $D_H$ from Tagawa et al.[4] and $D_C$ from Fischer et al.[5] Red and blue regions in **b** indicate $H_2O$ and C concentrations in non-carbonaceous and carbonaceous chondrites, respectively. See Supplementary Table 8 for the $H_2O$ and C contents in each type of chondrites.

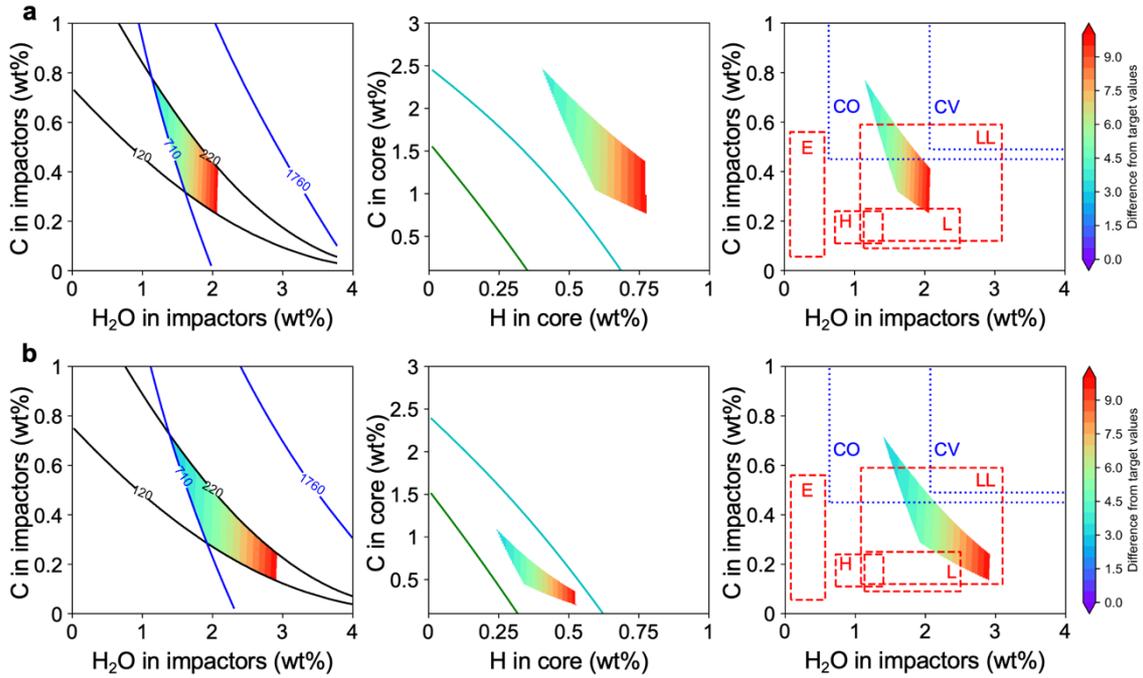

**Fig. 6 Modelling of homogeneous accretion & multi-stage core formation (model S1).** The delivery of water and carbon to the Earth is considered from the beginning (**a**) and only after 50% accretion (**b**). The colored region in the left panels indicates $H_2O$ and C concentrations in impactors explaining the BSE abundances of 710–1760 ppm $H_2O$ and 120–220 ppm C in addition to the core mass fraction and the mantle FeO, Ni and Co contents (color indicates the deviation from target values). Those in the central panels show the corresponding H and C contents in the core, which are compared with H and C concentrations required to explain the outer core density deficit when $T_{ICB}$ = 5400 K (green curve) and 4800 K (right blue curve) (see text). The $H_2O$ and C abundances in impactors (= Earth building blocks) are compared to those in non-carbonaceous (red region) and carbonaceous chondrites (blue region) in the right panels.

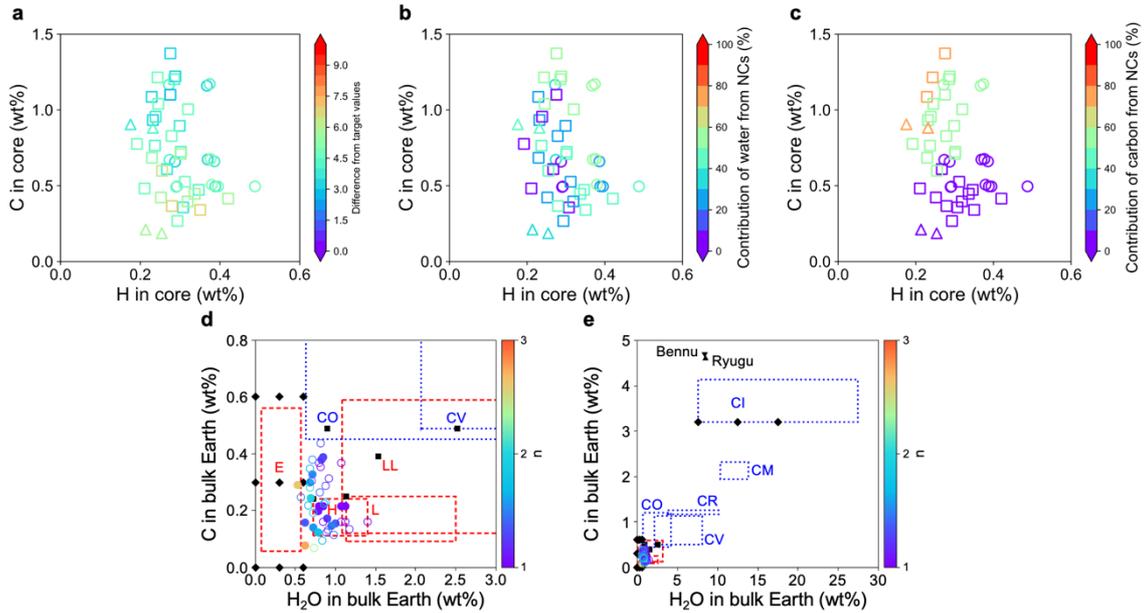

**Fig. 7 Modelling of heterogeneous accretion & multi-stage core formation.** Water and carbon are delivered according to the 4-stage accretion scenario by Dauphas et al.[46] that explains Earth's isotopic composition. **a–c** Modelling results on core H and C concentrations (circles, n =1; squares, 1 < n < 2; triangles 2 < n < 3). Colors indicate difference from target values (σ) (**a**) and the fraction of NC-type materials as sources for Earth's water (**b**) and carbon (**c**). **d, e** The bulk Earth $H_2O$ (present as H in the core) and C abundances compared to those in each type of chondrite (Supplementary Table 8) including the Ryugu[71] and Bennu[72] samples. The dependence on n is indicated by color. Closed and open symbols consider none and 25% loss of carbon during Earth accretion, respectively. Diamonds and squares show the variations in $H_2O$ and C concentrations in E-/CI-type materials (varied in each model) and other types (fixed), respectively, employed in our modelling, while the reported ranges of the $H_2O$ and C contents for NC and CC materials are indicated by boxes with dotted and broken lines, respectively.

Supplementary Information for

**Core-mantle partitioning and the bulk Earth abundances of hydrogen and carbon: Implications for their origins**

by Tsutsumi et al.

**Supplementary Table 1 Conditions and results of single-stage core formation modelling.**

| $P$ (GPa) | $T$ (K) | Ref. | $D_H^{metal/silicate}$ | $D_C^{metal/silicate}$ | Core composition (wt%) | | | | | Bulk Earth abundance (wt%) | |
|---|---|---|---|---|---|---|---|---|---|---|---|
| | | | | | H | C | O | Si | S | $H_2O$ | C |
| 40 | 3750 | Wade and Wood[18] | 92 | 67 | 0.68 | 0.75 | 0.00 | 3.43 | 2.00 | 2.08 | 0.26 |
| 50 | 3500 | Siebert et al.[19] | 56 | 96 | 0.45 | 1.15 | 2.52 | 4.42 | 2.00 | 1.38 | 0.39 |
| 54 | 3350 | Fischer et al.[20] | 44 | 107 | 0.35 | 1.29 | 1.49 | 7.78 | 2.00 | 1.10 | 0.44 |

$D_H$ and $D_C$ are based on molar concentratio

**Supplementary Table 2** Examples of parameter sets and results of homogeneous accretion & multi-stage core formation modelling when impactors include water and carbon from the beginning of Earth accretion.

| Paramerter sets | | | | | | | | | | | |
|---|---|---|---|---|---|---|---|---|---|---|---|
| Model # | | S1 | S2 | S3 | F1 | F2 | F3 | R1 | R2 | R3 | |
| $N$ | | 3 | 3 | 2 | 1 | 3 | 3 | 1 | 3 | 3 | |
| $P_{final}$ | | 54 | 62 | 64 | 48 | 62 | 62 | 44 | 46 | 46 | |
| $H_2O$ (wt%) | | 1.5 | 2.0 | 2.0 | 0.6 | 1.0 | 1.0 | 0.5 | 2.0 | 1.9 | |
| C (wt%) | | 0.5 | 0.5 | 0.5 | 0.5 | 0.5 | 0.5 | 0.5 | 0.5 | 0.5 | |
| Delivery of $H_2O$ and C | | 1–1000 steps | 1–1000 steps | 1–1000 steps | 1–1000 steps | 1–1000 steps | 1–1000 steps | 1–1000 steps | 1–1000 steps | 1–1000 steps | |
| Impactor core mass fraction | | 0.326 | 0.324 | 0.324 | 0.329 | 0.328 | 0.328 | 0.329 | 0.324 | 0.325 | |
| Impactor core radius (km) | | 274.4 | 274.0 | 313.6 | 396.9 | 274.9 | 274.9 | 397.0 | 274.0 | 274.0 | |
| Impactor silicate (wt%) | FeO | 0.73 | 0.73 | 0.73 | 0.74 | 0.74 | 0.74 | 0.74 | 0.73 | 0.73 | |
| | $SiO_2$ | 51.82 | 51.43 | 51.43 | 52.52 | 52.21 | 52.21 | 52.60 | 51.43 | 51.50 | |
| | MgO | 37.07 | 36.79 | 36.79 | 37.58 | 37.35 | 37.35 | 37.63 | 36.79 | 36.85 | |
| | $Al_2O_3$ | 4.52 | 4.48 | 4.48 | 4.58 | 4.55 | 4.55 | 4.58 | 4.48 | 4.49 | |
| | CaO | 3.64 | 3.61 | 3.61 | 3.69 | 3.67 | 3.67 | 3.69 | 3.61 | 3.62 | |
| | $H_2O$ | 2.23 | 2.96 | 2.96 | 0.89 | 1.49 | 1.49 | 0.75 | 2.96 | 2.81 | |
| Impactor core (wt%) | Fe | 89.18 | 89.16 | 89.16 | 89.21 | 89.20 | 89.20 | 89.21 | 89.16 | 89.17 | |
| | Ni | 5.28 | 5.28 | 5.28 | 5.28 | 5.28 | 5.28 | 5.28 | 5.28 | 5.28 | |
| | Co | 0.25 | 0.25 | 0.25 | 0.25 | 0.25 | 0.25 | 0.25 | 0.25 | 0.25 | |
| | O | 0.00 | 0.00 | 0.00 | 0.00 | 0.00 | 0.00 | 0.00 | 0.00 | 0.00 | |
| | Si | 1.76 | 1.76 | 1.76 | 1.76 | 1.76 | 1.76 | 1.76 | 1.76 | 1.76 | |
| | S | 1.99 | 2.00 | 2.00 | 1.98 | 1.98 | 1.98 | 1.97 | 2.00 | 2.00 | |
| | C | 1.53 | 1.54 | 1.54 | 1.52 | 1.53 | 1.53 | 1.52 | 1.54 | 1.54 | |
| Results | | | | | | | | | | | Present Earth |
| Core mass fraction | | 0.299 | 0.289 | 0.289 | 0.300 | 0.294 | 0.294 | 0.302 | 0.283 | 0.286 | 0.325 |
| Bulk silicate Earth (wt%) | FeO | 8.57 | 7.94 | 8.01 | 8.71 | 8.44 | 8.40 | 8.31 | 8.13 | 7.48 | 8.13 |
| | $SiO_2$ | 47.65 | 48.96 | 48.88 | 47.07 | 47.82 | 47.85 | 47.26 | 49.16 | 49.59 | |
| | MgO | 35.61 | 34.96 | 34.96 | 36.02 | 35.54 | 35.55 | 36.17 | 34.66 | 34.86 | |
| | $Al_2O_3$ | 4.34 | 4.26 | 4.26 | 4.39 | 4.33 | 4.33 | 4.41 | 4.22 | 4.25 | |
| | CaO | 3.50 | 3.43 | 3.43 | 3.54 | 3.49 | 3.49 | 3.55 | 3.40 | 3.42 | |
| | CoO (ppm) | 127 | 138 | 140 | 94 | 128 | 129 | 152 | 163 | 150 | 131 |
| | NiO (ppm) | 2266 | 2528 | 2610 | 1465 | 2640 | 2658 | 1470 | 2110 | 1999 | 2418 |
| | $H_2O$ (ppm) | 766 | 1665 | 1674 | 985 | 856 | 846 | 1190 | 1756 | 1681 | 710–1760 |
| | C (ppm) | 175 | 181 | 187 | 147 | 141 | 144 | 135 | 203 | 164 | 120–220 |
| Earth core (wt%) | Fe | 83.06 | 86.21 | 86.09 | 83.31 | 85.11 | 85.16 | 83.58 | 87.61 | 87.99 | |
| | Ni | 5.35 | 5.44 | 5.42 | 5.53 | 5.40 | 5.40 | 5.49 | 5.64 | 5.60 | |
| | Co | 0.25 | 0.25 | 0.25 | 0.26 | 0.25 | 0.25 | 0.24 | 0.25 | 0.25 | |
| | O | 2.73 | 1.56 | 1.61 | 1.41 | 0.94 | 0.92 | 1.28 | 0.58 | 0.49 | |
| | Si | 4.26 | 1.88 | 1.96 | 5.50 | 4.06 | 4.03 | 5.47 | 1.17 | 0.99 | |
| | S | 2.18 | 2.25 | 2.25 | 2.17 | 2.21 | 2.21 | 2.15 | 2.30 | 2.27 | |
| | H | 0.54 | 0.73 | 0.73 | 0.20 | 0.36 | 0.36 | 0.15 | 0.74 | 0.70 | |
| | C | 1.63 | 1.68 | 1.68 | 1.63 | 1.67 | 1.67 | 1.62 | 1.72 | 1.71 | |
| $H_2O$ in Bulk Earth (wt%) | | 1.5 | 2.0 | 2.0 | 0.6 | 1.0 | 1.0 | 0.5 | 2.0 | 1.9 | |
| C in Bulk Earth (wt%) | | 0.5 | 0.5 | 0.5 | 0.5 | 0.5 | 0.5 | 0.5 | 0.5 | 0.5 | |
| Difference from target values σ | | 5.31 | 6.16 | 6.17 | 8.16 | 5.71 | 5.66 | 6.46 | 7.85 | 8.19 | |

Supplementary Table 3  Examples of parameter sets and results of homogeneous accretion & multi-stage core formation modelling when impactors include water and carbon after 50% Earth accretion.

| Paramerter sets | | | | | | | | | | |
|---|---|---|---|---|---|---|---|---|---|---|
| Model # | | S1 | S2 | S3 | F1 | F2 | F3 | R1 | R2 | R3 |
| $N$ | | 3 | 3 | 2 | 1 | 3 | 3 | 1 | 3 | 3 |
| $P_{final}$ | | 56 | 60 | 62 | 42 | 64 | 62 | 42 | 42 | 42 |
| $H_2O$ (wt%) | | 2.00 | 2.30 | 2.30 | 2.00 | 1.50 | 1.50 | 2.00 | 2.90 | 3.20 |
| C (wt%) | | 0.40 | 0.40 | 0.40 | 0.40 | 0.40 | 0.40 | 0.40 | 0.20 | 0.15 |
| Delivery of $H_2O$ and C | | 501–1000 steps | 501–1000 steps | 501–1000 steps | 501–1000 steps | 501–1000 steps | 501–1000 steps | 501–1000 steps | 501–1000 steps | 501–1000 steps |
| Impactor core mass fraction | | 0.324 | 0.323 | 0.323 | 0.324 | 0.325 | 0.325 | 0.324 | 0.320 | 0.318 |
| Impactor core radius (km) | | 273.8 | 273.5 | 313.1 | 394.8 | 274.2 | 274.2 | 394.8 | 272.6 | 272.2 |
| Impactor silicate (wt%) | FeO | 0.73 | 0.72 | 0.72 | 0.73 | 0.73 | 0.73 | 0.73 | 0.72 | 0.71 |
| | $SiO_2$ | 51.43 | 51.20 | 51.20 | 51.43 | 51.82 | 51.82 | 51.43 | 50.74 | 50.51 |
| | MgO | 36.79 | 36.63 | 36.63 | 36.79 | 37.07 | 37.07 | 36.79 | 36.30 | 36.14 |
| | $Al_2O_3$ | 4.48 | 4.46 | 4.46 | 4.48 | 4.52 | 4.52 | 4.48 | 4.42 | 4.40 |
| | CaO | 3.61 | 3.60 | 3.60 | 3.61 | 3.64 | 3.64 | 3.61 | 3.56 | 3.55 |
| | $H_2O$ | 2.96 | 3.40 | 3.40 | 2.96 | 2.22 | 2.22 | 2.96 | 4.26 | 4.69 |
| Impactor core (wt%) | Fe | 89.44 | 89.43 | 89.43 | 89.44 | 89.46 | 89.46 | 89.44 | 89.98 | 90.12 |
| | Ni | 5.30 | 5.30 | 5.30 | 5.30 | 5.30 | 5.30 | 5.30 | 5.33 | 5.34 |
| | Co | 0.25 | 0.25 | 0.25 | 0.25 | 0.25 | 0.25 | 0.25 | 0.25 | 0.25 |
| | O | 0.00 | 0.00 | 0.00 | 0.00 | 0.00 | 0.00 | 0.00 | 0.00 | 0.00 |
| | Si | 1.77 | 1.77 | 1.77 | 1.77 | 1.77 | 1.77 | 1.77 | 1.78 | 1.78 |
| | S | 2.01 | 2.01 | 2.01 | 2.01 | 2.00 | 2.00 | 2.01 | 2.03 | 2.04 |
| | C | 1.24 | 1.24 | 1.24 | 1.24 | 1.23 | 1.23 | 1.24 | 0.63 | 0.47 |
| Results | | | | | | | | | | | Present Earth |
| Core mass fraction | | 0.301 | 0.297 | 0.298 | 0.295 | 0.293 | 0.294 | 0.294 | 0.289 | 0.286 | 0.325 |
| Bulk silicate Earth (wt%) | FeO | 8.59 | 7.11 | 6.99 | 8.02 | 8.73 | 8.47 | 8.54 | 7.32 | 7.37 | 8.13 |
| | $SiO_2$ | 47.07 | 48.82 | 48.90 | 48.08 | 47.26 | 47.51 | 47.64 | 49.31 | 49.41 | |
| | MgO | 36.04 | 35.79 | 35.83 | 35.75 | 35.71 | 35.75 | 35.65 | 35.27 | 35.12 | |
| | $Al_2O_3$ | 4.39 | 4.36 | 4.36 | 4.35 | 4.35 | 4.35 | 4.34 | 4.30 | 4.28 | |
| | CaO | 3.54 | 3.51 | 3.52 | 3.51 | 3.51 | 3.51 | 3.50 | 3.46 | 3.45 | |
| | CoO (ppm) | 136 | 124 | 124 | 79 | 141 | 133 | 153 | 143 | 146 | 131 |
| | NiO (ppm) | 2534 | 2245 | 2305 | 1149 | 2991 | 2751 | 1479 | 1787 | 1918 | 2418 |
| | $H_2O$ (ppm) | 824 | 1542 | 1455 | 1421 | 1165 | 1104 | 1413 | 1277 | 1486 | 710–1760 |
| | C (ppm) | 176 | 153 | 152 | 167 | 151 | 148 | 190 | 166 | 198 | 120–220 |
| Earth core (wt%) | Fe | 83.15 | 86.71 | 86.73 | 85.68 | 85.31 | 85.60 | 85.23 | 88.57 | 88.98 | |
| | Ni | 5.31 | 5.42 | 5.39 | 5.66 | 5.38 | 5.41 | 5.64 | 5.65 | 5.66 | |
| | Co | 0.25 | 0.25 | 0.25 | 0.26 | 0.25 | 0.25 | 0.25 | 0.25 | 0.26 | |
| | O | 2.66 | 1.14 | 1.19 | 1.15 | 0.96 | 0.88 | 1.25 | 0.44 | 0.39 | |
| | Si | 5.51 | 3.27 | 3.24 | 4.06 | 4.98 | 4.74 | 4.44 | 2.00 | 1.64 | |
| | S | 2.16 | 2.19 | 2.18 | 2.20 | 2.22 | 2.21 | 2.21 | 2.25 | 2.27 | |
| | H | 0.35 | 0.39 | 0.39 | 0.34 | 0.26 | 0.26 | 0.34 | 0.53 | 0.58 | |
| | C | 0.62 | 0.64 | 0.64 | 0.64 | 0.65 | 0.65 | 0.64 | 0.31 | 0.21 | |
| $H_2O$ in Bulk Earth (wt%) | | 1.00 | 1.15 | 1.15 | 1.00 | 1.25 | 1.25 | 1.00 | 1.45 | 1.60 | |
| C in Bulk Earth (wt%) | | 0.20 | 0.20 | 0.20 | 0.20 | 0.20 | 0.20 | 0.20 | 0.10 | 0.075 | |
| Difference from target values σ | | 5.03 | 8.16 | 8.71 | 9.45 | 7.24 | 5.90 | 7.77 | 8.68 | 8.60 | |

**Supplementary Table 4 Compositions of silicate melt and silicate starting materials (wt%).**

| | SiO$_2$ | TiO$_2$ | Al$_2$O$_3$ | MgO | FeO | CaO | Na$_2$O | K$_2$O | H$_2$O (SIMS) | C (SIMS) (ppm) | Total |
|---|---|---|---|---|---|---|---|---|---|---|---|
| Run # | | | | | | | | | | | |
| 1 | 29.84 (49) | 2.66 (14) | 9.56 (32) | 11.13 (11) | 25.25 (160) | 5.37 (6) | 1.67 (36) | 0.39 (6) | 10.78 (383) | 6095 (1413) | 97.25 |
| 2 | 36.21 (54) | 1.09 (3) | 12.16 (13) | 16.34 (26) | 21.83 (127) | 4.07 (10) | 3.37 (31) | 0.07 (1) | 0.11 (2) | 229 (90) | 95.27 |
| 3 | 39.17 (27) | 1.62 (8) | 11.57 (5) | 11.78 (11) | 25.16 (72) | 3.49 (6) | 7.37 (86) | 0.39 (5) | 1.19 (11) | 1197 (294) | 101.85 |
| 4 | 41.01 (24) | 2.34 (6) | 22.51 (17) | 11.16 (9) | 15.51 (89) | 3.36 (11) | 3.42 (44) | 0.04 (1) | 1.49 (42) | 1836 (1199) | 101.04 |
| 5 | 39.96 (113) | 2.43 (10) | 22.73 (10) | 11.28 (16) | 14.76 (137) | 3.48 (14) | 2.92 (47) | 0.04 (1) | 2.50 (57) | 2127 (401) | 100.32 |
| 6 | 38.68 (42) | 1.72 (6) | 10.99 (20) | 10.00 (12) | 25.00 (35) | 4.48 (7) | 4.67 (20) | 0.13 (2) | - | 354 (101) | 95.71 |
| 7 | 26.73 (60) | 5.23 (39) | 11.55 (61) | 6.13 (35) | 35.06 (79) | 2.35 (10) | 5.05 (37) | 0.11 (4) | - | 1218 (545) | 92.33 |
| 8 | 38.93 (70) | 1.45 (3) | 16.75 (47) | 8.33 (25) | 24.66 (36) | 4.13 (12) | 2.65 (11) | 0.22 (3) | - | 257 (113) | 97.15 |
| 9 | 43.96 (173) | 1.84 (16) | 26.38 #### | 12.27 (60) | 9.79 (144) | 4.74 (39) | 3.40 (26) | 0.33 (3) | - | 81 (71) | 102.72 |
| 10 | 24.55 (35) | 5.06 (17) | 9.99 (31) | 3.84 (9) | 46.00 (100) | 1.81 (5) | 4.64 (17) | 0.21 (2) | - | 4510 (1204) | 96.55 |
| Silicate starting materials | | | | | | | | | | | |
| w01[a] | 49.64 | 1.64 | 14.88 | 8.51 | 11.43 | 10.55 | 2.90 | 0.12 | 1.00 | - | 99.67 |
| w02 | 47.67 (33) | 1.50 (8) | 6.66 (9) | 12.19 (4) | 8.19 (17) | 9.84 (6) | 2.19 (9) | 0.11 (1) | 0.15 | - | 88.35 |
| w03 | 43.58 (65) | 1.38 (9) | 13.45 (11) | 15.96 (15) | 5.99 (10) | 8.84 (9) | 3.65 (14) | 0.03 (1) | 0.60 | - | 92.87 |
| n01[b] | 47.60 | 4.28 | 14.43 | 8.29 | 8.76 | 9.43 | 2.62 | 0.08 | - | - | 95.61 |
| n02 | 48.59 (57) | 1.00 (6) | 15.20 (11) | 8.57 (7) | 9.40 (35) | 9.36 (14) | 3.40 (11) | 0.14 (2) | - | - | 95.66 |

[a]Same as the starting material used in Tagawa et al.[4].

[b]0.12 wt% MnO was included.

**Supplementary Table 5 3rd-order Birch-Murnaghan equations of state for $(FeH_x)_7C_3$.**

| | hydrogen concentration | $V_0$ (Å$^3$/unit-cell) | $V_0$ (Å$^3$/Fe atom) | $K_0$ (GPa) | $K'_0$ |
|---|---|---|---|---|---|
| $Fe_7C_3$ | | 181.40 | 12.96 | 254.5 | 3.2 |
| $Fe_7C_3H$ (2a) | $x = 0.14$ | 184.67 | 13.19 | 240.8 | 4.1 |
| $Fe_7C_3H_2$ (2a+2b) | $x = 0.29$ | 189.61 | 13.54 | 215.5 | 5.3 |
| $Fe_7C_3H_4$ (2a+6c) | $x = 0.57$ | 197.92 | 14.14 | 242.2 | 4.4 |

**Supplementary Table 6 Hydrogen concentrations $x$ in quench crystals formed from liquid metal.**

| Run # | Quench crystals | $x$ in fcc FeH$_x$ | $x$ in (FeH$_x$)$_7$C$_3$ | $x$ in (FeH$_x$)$_3$C | H/(Fe+Si) in liquid |
|---|---|---|---|---|---|
| 1 | fcc-FeH$_x$ + (FeH$_x$)$_7$C$_3$ + ε-FeOOH | 1.57 (13) | 0.29 (2) |  | 0.30 (3) |
| 2 | fcc-FeH$_x$ + (FeH$_x$)$_3$C + ε-FeOOH | 0.03 (1) |  | 0.22 (5) | 0.20 (3) |
| 3 | fcc-FeH$_x$ + (FeH$_x$)$_7$C$_3$ + ε-FeOOH | 1.76 (11) | 0.21 (4) |  | 0.64 (6) |
| 4 | fcc-FeH$_x$ + (FeH$_x$)$_7$C$_3$ + ε-FeOOH | 1.67 (3) | 0.30 (6) |  | 0.61 (6) |
| 5 | fcc-FeH$_x$ + (FeH$_x$)$_3$C + ε-FeOOH | 1.51 (7) |  | 0.45 (4) | 0.50 (4) |

**Supplementary Table 7  Compositions of quenched liquid metal (wt%).**

| Run # | Fe | Al | Ca | Ti | Si | O | C | H (XRD) | Total |
|---|---|---|---|---|---|---|---|---|---|
| 1 | 84.80 (323) | 0.01 (1) | 0.07 (2) | 0.07 (4) | 0.17 (9) | 0.96 (49) | 9.51 (191) | 0.47 (5) | 95.59 |
| 2 | 81.57 (29) | 0.11 (1) | 0.04 (1) | 0.10 (2) | 3.70 (18) | 4.23 (20) | 3.20 (20) | 0.33 (4) | 92.95 |
| 3 | 83.87 (70) | 0.02 (1) | 0.07 (2) | 0.07 (3) | 0.52 (3) | 4.19 (9) | 5.32 (11) | 0.98 (7) | 94.06 |
| 4 | 83.39 (160) | 0.12 (2) | 0.04 (1) | 0.19 (3) | 2.33 (10) | 3.82 (54) | 5.98 (6) | 0.97 (8) | 95.86 |
| 5 | 77.78 (65) | 0.30 (1) | 0.07 (1) | 0.38 (4) | 3.04 (12) | 4.48 (23) | 7.61 (21) | 0.76 (5) | 93.67 |
| 6 | 86.07 (107) | 0.06 (2) | 0.06 (2) | 0.08 (3) | 1.06 (12) | 3.29 (35) | 3.11 (20) | | 93.73 |
| 7 | 76.79 (162) | 0.13 (1) | 0.14 (2) | 0.15 (2) | 0.28 (2) | 2.20 (30) | 9.10 (52) | | 88.79 |
| 8 | 90.31 (31) | 0.04 (2) | 0.06 (1) | 0.08 (2) | 1.31 (23) | 3.49 (44) | 3.99 (9) | | 99.29 |
| 9 | 86.10 (73) | 0.11 (2) | 0.04 (1) | 0.12 (3) | 6.47 (8) | 2.62 (12) | 3.04 (10) | | 98.49 |
| 10 | 79.76 (88) | 0.07 (1) | 0.07 (1) | 0.31 (2) | 0.16 (2) | 6.11 (8) | 7.00 (9) | | 93.48 |

**Supplementary Table 8  Water and carbon concentrations in each chondrite, Ryugu and Bennu.**

| $H_2O$ (wt%) | Alexander[41,42] | Alexander[a,41,42] | Lodders[38] | Piani et al.[2] | Wasson[39] | Yokoyama et al.[71] | Lauretta et al.[72] |
|---|---|---|---|---|---|---|---|
| CI | 13.95 | 27.50 | 16.74 | 7.54–9.08 | 18.00 | | |
| CM | 10.35 | 13.80 | 10.35 | | 12.60 | | |
| CR | 4.41 | 10.10 | 3.78 | | | | |
| CO | 4.14 | 2.70 | 0.90 | | 0.63 | | |
| CV | 2.07 | 8.10 | 2.52 | | 2.52 | | |
| H | | 1.40 | 0.72 | | | | |
| L | | 2.50 | 1.13 | | | | |
| LL | 1.08 | 3.10 | 1.53 | | | | |
| EH | | | 0.57 | 0.19–0.54 | | | |
| EL | | | | 0.08–0.41 | | | |
| Ryugu | | | | | | 8.46 | |
| Bennu | | | | | | | 8.37 |

| C (wt%) | Alexander[41,42] | Lodders[38] | Wasson[39] | Moore & Rewis[40] | Yokoyama et al.[71] | Lauretta et al.[72] |
|---|---|---|---|---|---|---|
| CI | 3.65 | 4.13 | 3.20 | | | |
| CM | 1.94 | 2.32 | 2.20 | | | |
| CR | 1.17 | 1.26 | | | | |
| CO | 1.20 | 0.49 | 0.45 | | | |
| CV | 1.13 | 0.49 | 0.56 | | | |
| H | | 0.24 | 0.11 | | | |
| L | | 0.25 | 0.09 | | | |
| LL | 0.59 | 0.39 | 0.12 | | | |
| EH | 0.40 | 0.38 | 0.40 | 0.22–0.56 | | |
| EL | 0.43 | 0.36 | | 0.056–0.43 | | |
| Ryugu | | | | | 4.60 | |
| Bennu | | | | | | 4.70 |

[a]Estimated original $H_2O$ contents before oxidizing iron, considering that chondrites accreted their water as ice which subsequently drove aqueous alterations.

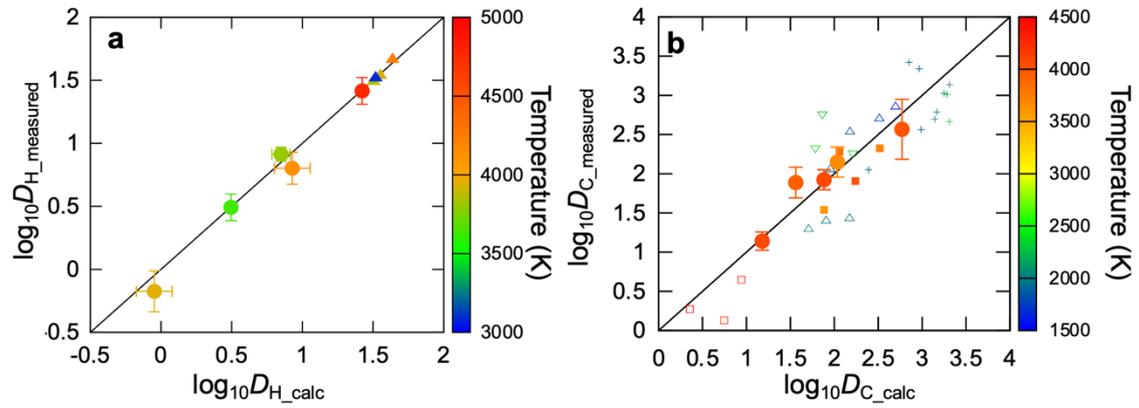

**Supplementary Fig. 1 Comparison between experimental data and fitting results. a** $D_H$ (Eq. 3) and **b** $D_C$ (Eq. 5). Closed circles, this study; closed triangles, Tagawa et al.[4]; squares, Blanchard et al.[6] (closed) and Fischer et al.[5] (open); triangles, Fichtner et al.[17] (normal) and Dasgupta et al.[13] (inverted); crosses, Kuwahara et al.[14,15]

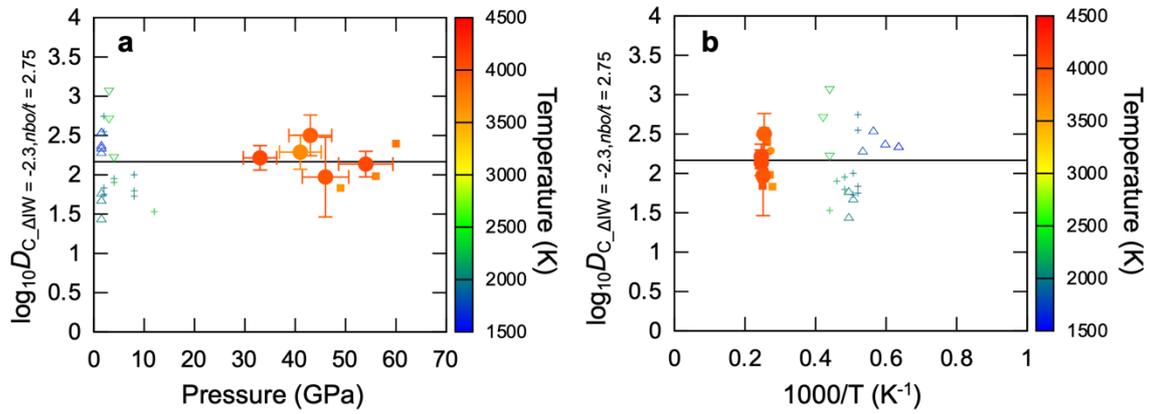

**Supplementary Fig. 2 Pressure and temperature dependence of $D_C$.** The measured $D_C$ values are adjusted to those at $\Delta$IW = -2.3 and $nbo/t$ = 2.75 using Eq. 5. Both pressure and temperature effects are found to be negligible and therefore not considered. Same symbols as those in Fig. 4 and Supplementary Fig. 1.

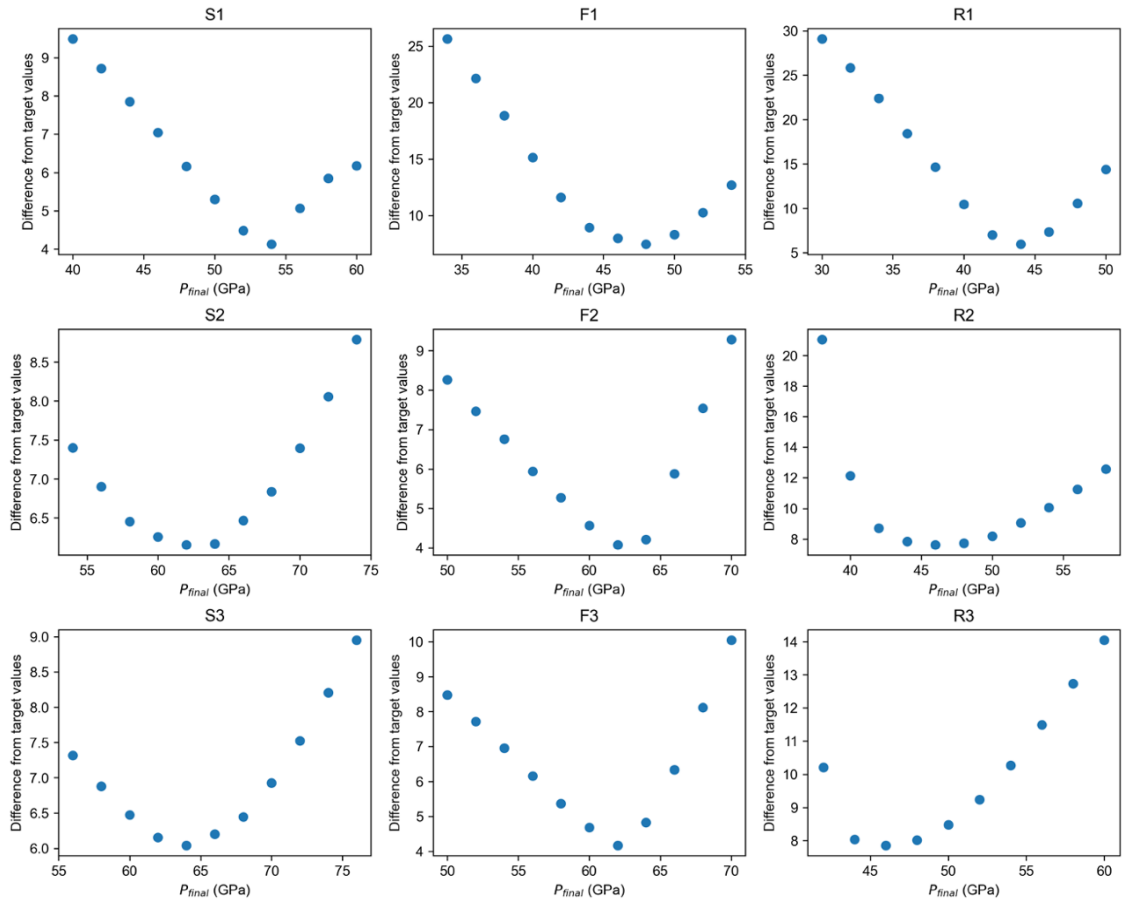

**Supplementary Fig. 3** Search for $P_{final}$ that minimizes the σ value (total difference from target values) in homogeneous accretion & multi-stage core formation modelling considering the delivery of water and carbon from the beginning of the Earth accretion.

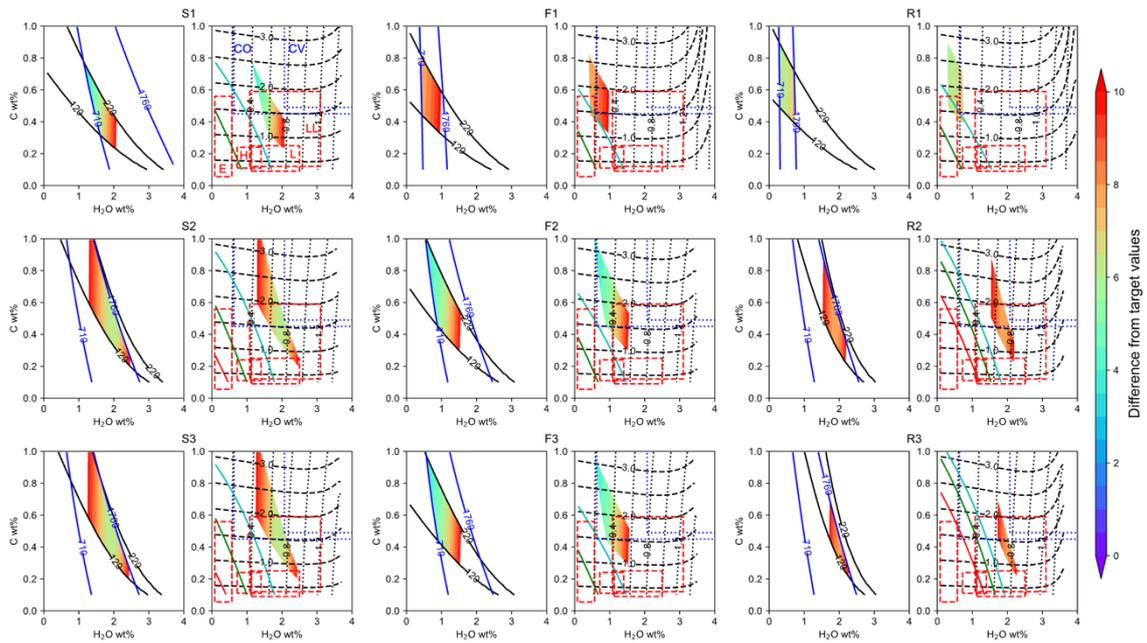

**Supplementary Fig. 4 Modelling of homogeneous accretion & multi-stage core formation (all models considering the delivery of water and carbon from the beginning of Earth accretion).** Similar to Fig. 6a except that calculated core concentrations of H and C (black dotted and broken lines, respectively) and those required to explain the outer core density deficit when $T_{ICB}$ = 6000 K (red curve), 5400 K (green) and 4800 K (right blue) are given in the right panel for each model. $H_2O$ and C concentrations in each non-carbonaceous (red) and carbonaceous chondrite (blue) are also shown. See text for more details.

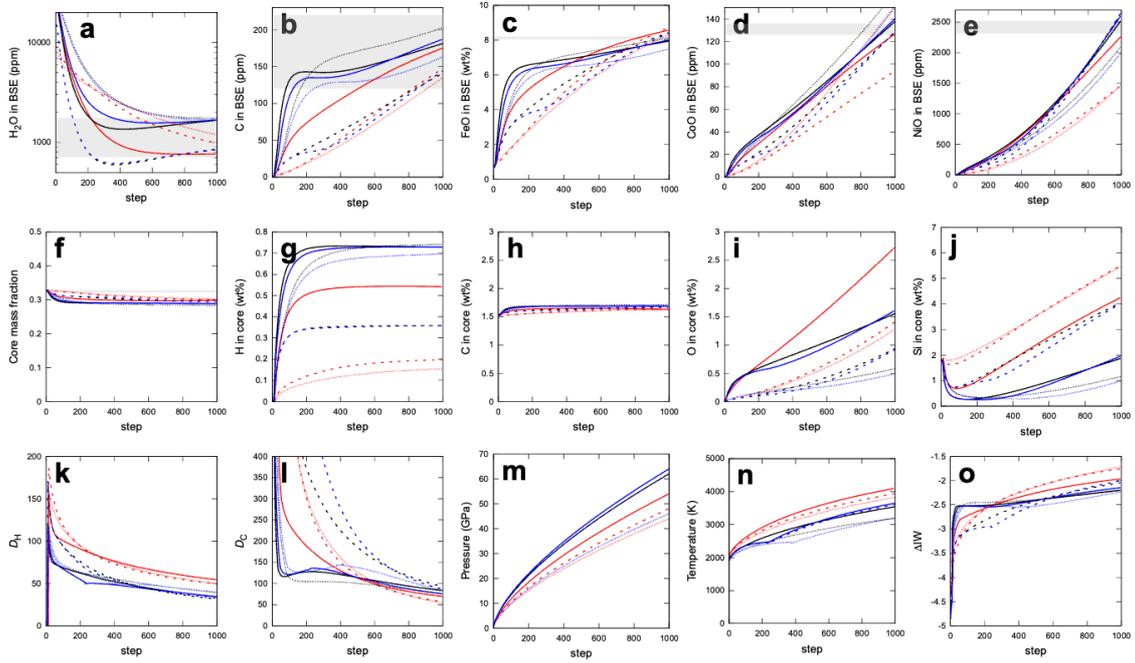

**Supplementary Fig. 5 Evolutions in homogeneous accretion & multi-stage core formation for all models (the delivery of water and carbon from the beginning of Earth accretion).** $H_2O$ (**a**), C (**b**), FeO (**c**), CoO (**d**) and NiO contents (**e**) in the BSE, core mass fraction (**f**), the hydrogen (**g**), carbon (**h**), oxygen (**i**) and silicon abundances (**j**) in the core, $D_H$ (**k**), $D_C$ (**l**), pressure (**m**), temperature (**n**) and ΔIW (**o**). Nine different $P\text{-}T$ evolutions of core formation are employed (solid curves, S1–S3; regular dotted curves, F1–F3; fine dotted curves, R1–R3). Colors indicate $P\text{-}T$ path (red, $T_{MOB1}$; black, $T_{MOB2}$; blue, $T_{MOB3}$). The gray bands in **a–f** show the BSE abundances and the core mass fraction (target values). Details are provided in Supplementary Table 2.

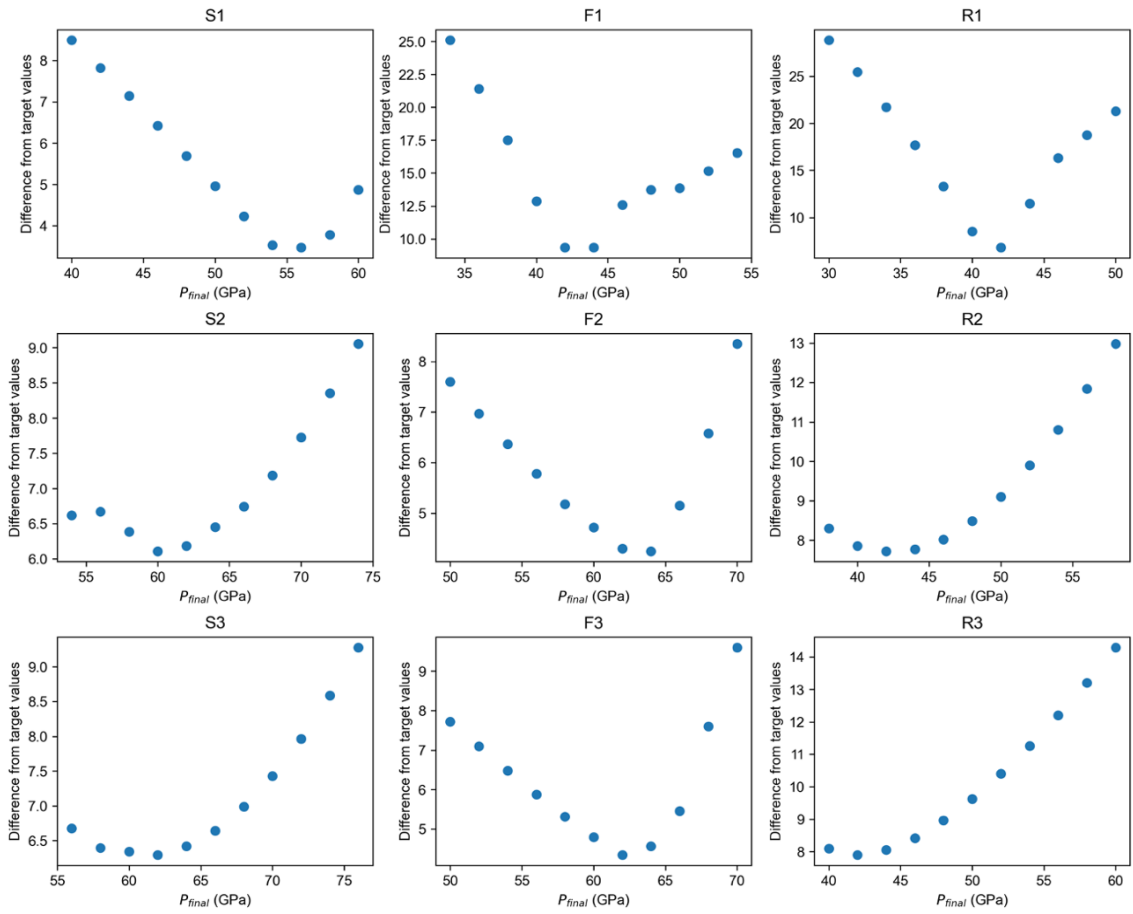

**Supplementary Fig. 6** Search for $P_{final}$ that minimizes the σ value (total difference from target values) in homogeneous accretion & multi-stage core formation modelling considering the delivery of water and carbon after 50% Earth accretion.

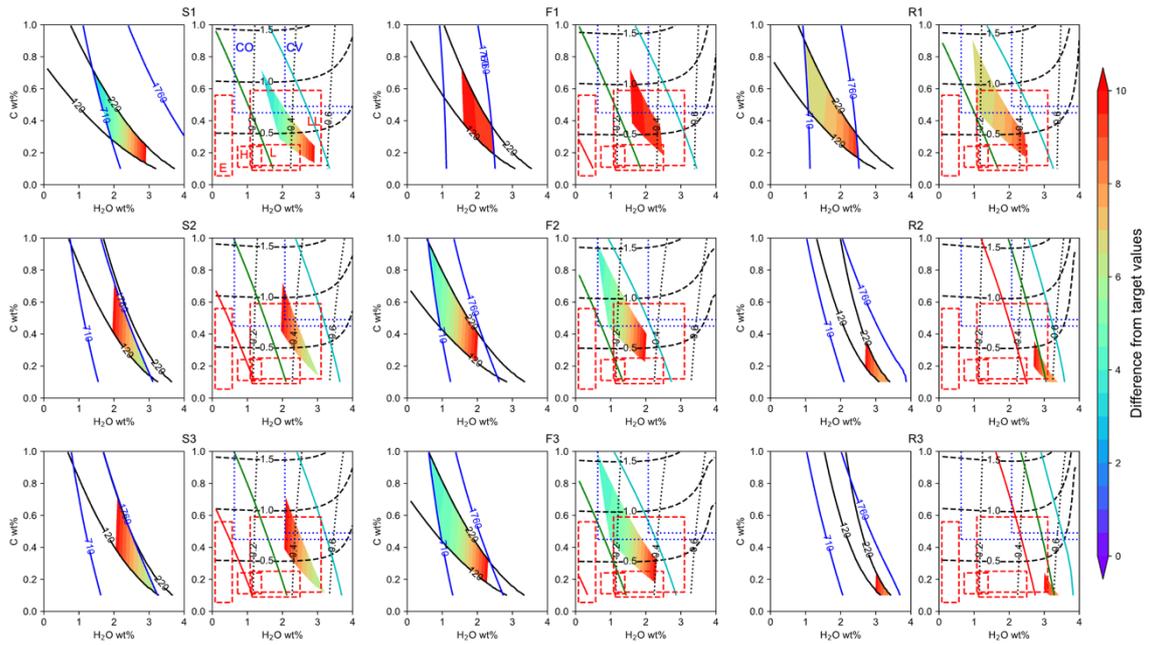

**Supplementary Fig. 7 Modelling of homogeneous accretion & multi-stage core formation (all models considering the delivery of water and carbon after 50% Earth accretion).** Similar to Supplementary Fig. 4. See text for more details.

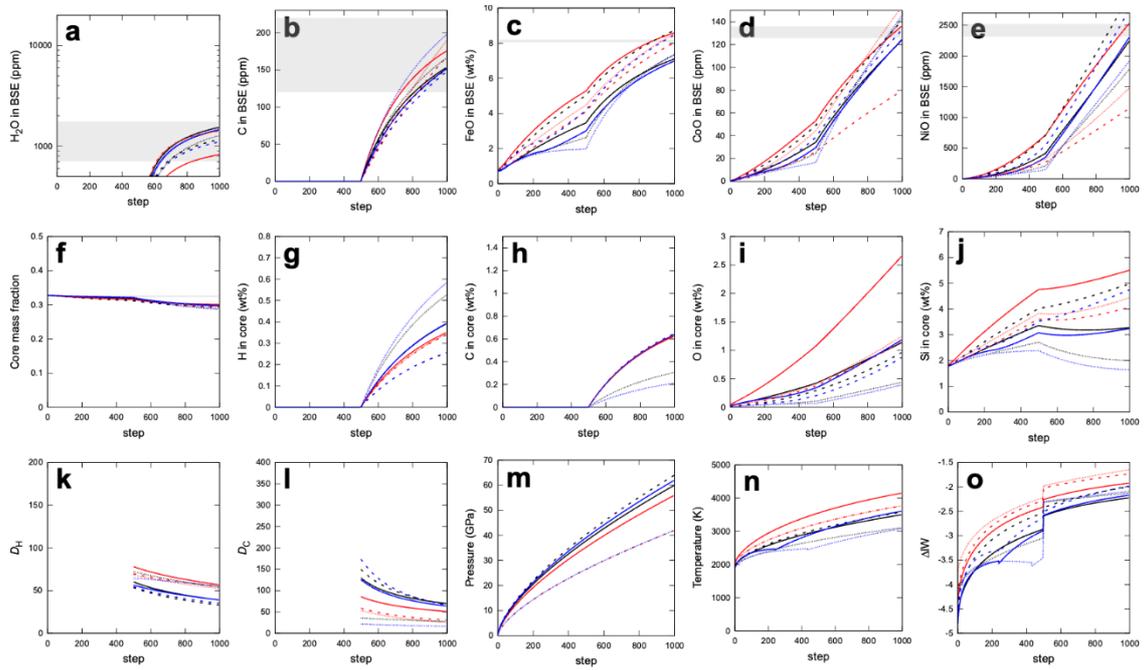

**Supplementary Fig. 8 Evolutions in homogeneous accretion & multi-stage core formation for all models (the delivery of water and carbon after 50% Earth accretion).** Similar to Supplementary Fig. 5. The gray bands in **a–f** show the BSE abundances and the core mass fraction (target values). Details are provided in Supplementary Table 3.

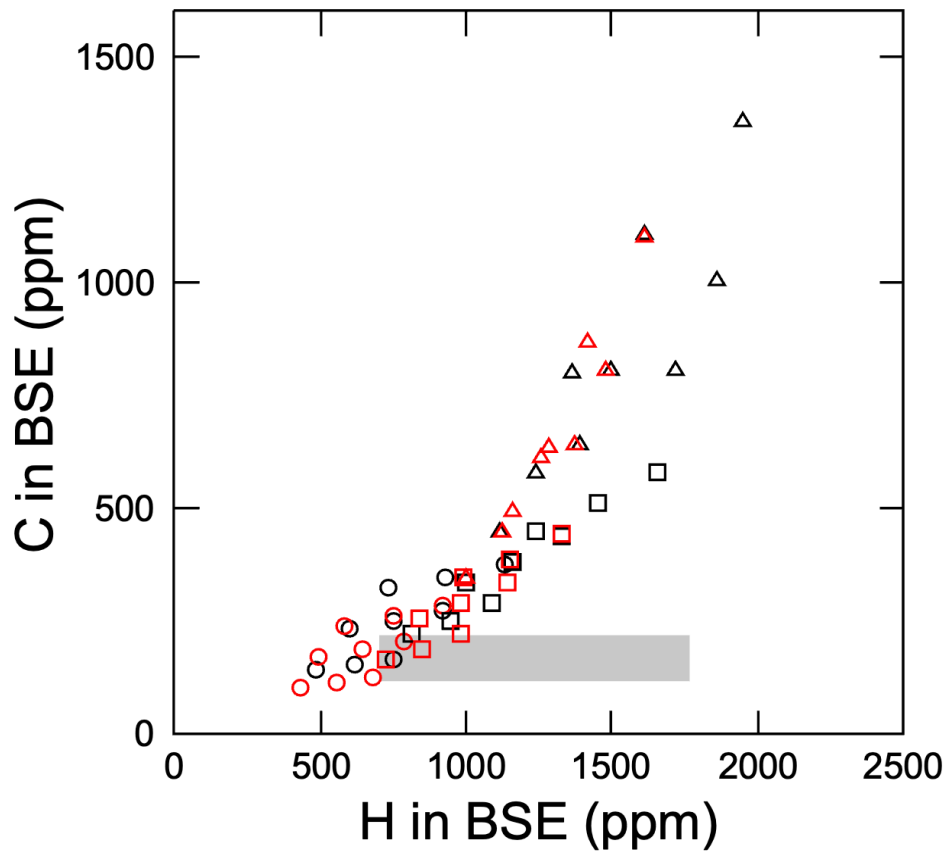

**Supplementary Fig. 9 Effects of H$_2$O and C concentrations in impactors on their BSE abundances obtained by heterogeneous accretion & multi-stage core formation modelling.** Here we employed 7.5 (circles), 12.5 (squares) and 17.5 wt% H$_2$O (triangles) and 3.2 wt% C in CI-type material (the C abundance was fixed since its effect is minor compared to that of the E-type composition) and 0, 0.3 and 0.6 wt% H$_2$O and 0, 0.3 and 0.6 wt% C in E-type one. Black and red symbols indicate none and 25% loss of carbon during Earth accretion, respectively. $N = 3$ and n = 1 were fixed in these calculations. A gray box shows the target ranges of H$_2$O and C concentrations in the BSE. Note that the calculated BSE H$_2$O and C contents form a narrow band by changing water and carbon abundances in E- and CI-type materials within reported values (Figs. 7c, d, Supplementary Table 8), resulting in a limited number of acceptable model conditions despite wide parameter ranges.

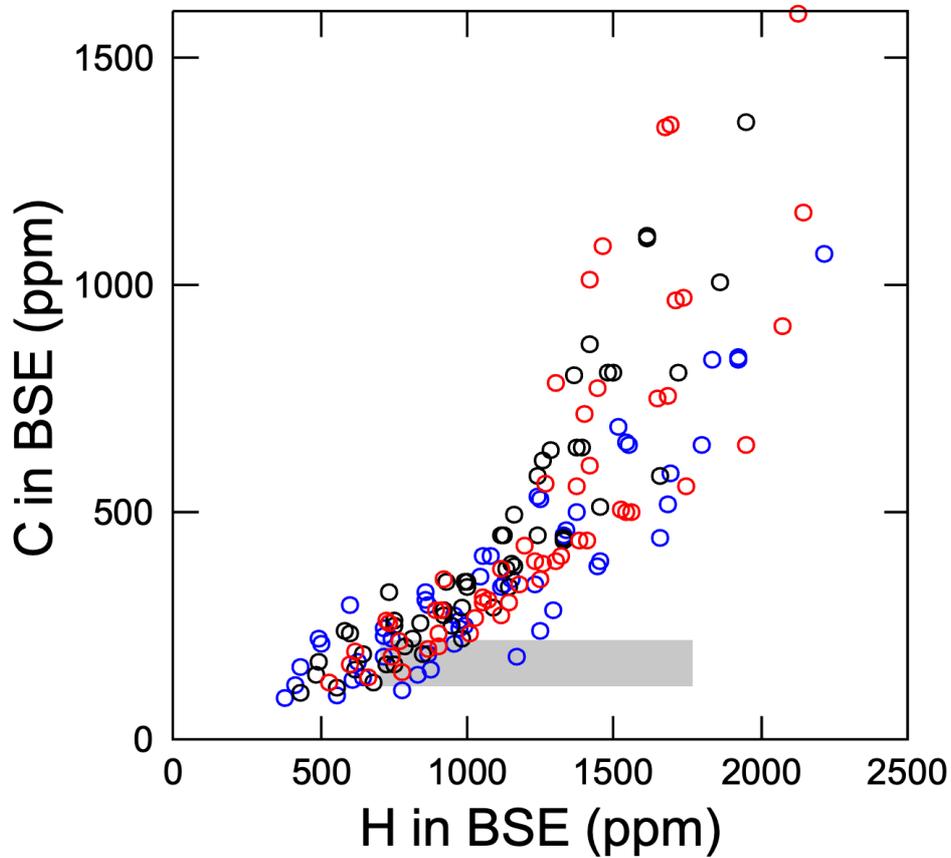

**Supplementary Fig. 10 Effect of impactor size on the BSE $H_2O$ and C contents obtained by heterogeneous accretion & multi-stage core formation modelling.** Similar to Supplementary Fig. 9 employing $N = 3$ (black symbols) but here $N$ is varied from 1 (blue) to 10 (red). $N = 10$ represents the impactor size corresponding to a geometric mean of the size of planetesimals given by Rubie et al.[3] Note that the effect of variations in $N$ (impactor size) on calculated BSE $H_2O$ and C abundances is limited.

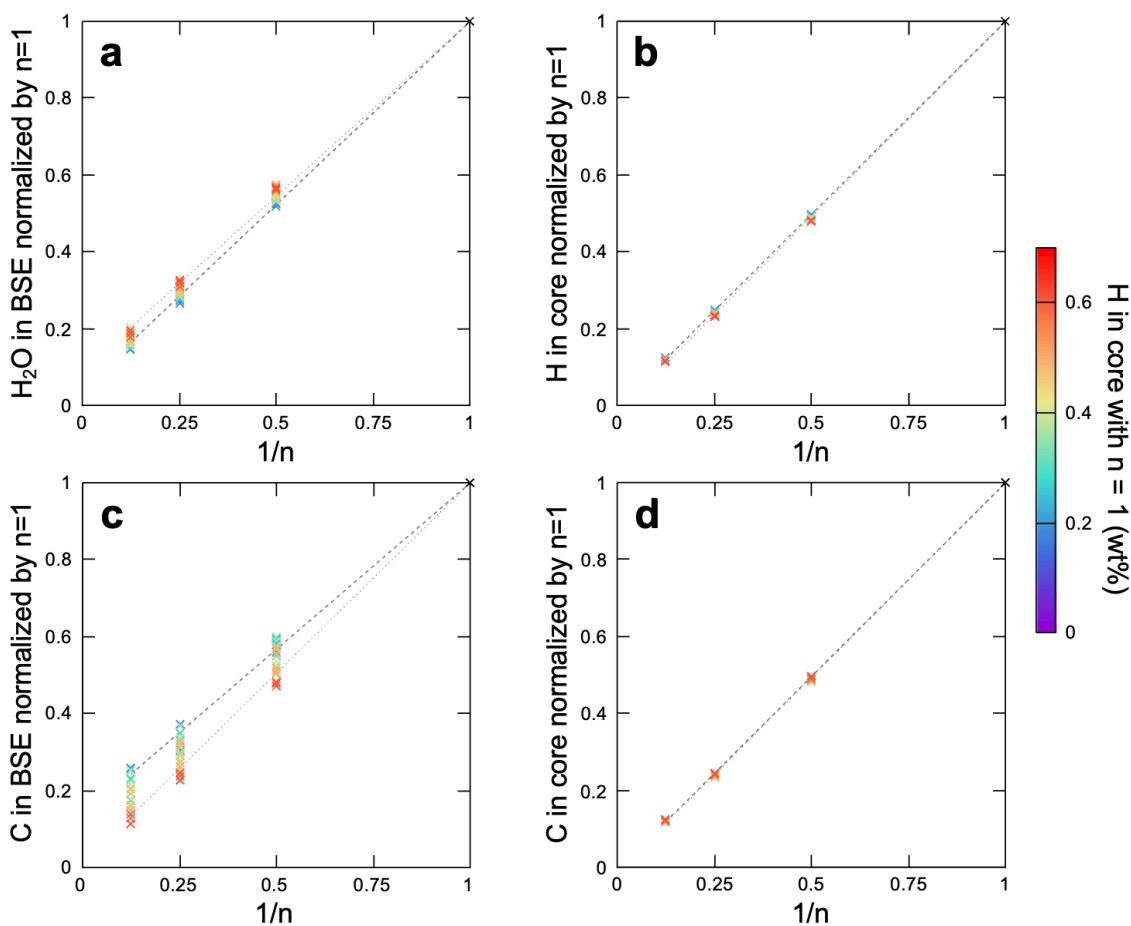

**Supplementary Fig. 11 Effect of differentiated impactors in heterogeneous accretion & multi-stage core formation modelling.** Here we consider volatile-free impactors with frequency of 1-1/n (1/n represents the frequency of $H_2O$ & C-bearing undifferentiated impactors) and calculated **a** BSE $H_2O$, **b** core H, **c** BSE C content and **d** core C contents by changing n = 1, 2, 4 and 8 (while fixing $N$ = 3) and the $H_2O$ and C contents in E- and CI-type materials within their reported wide ranges same as those in Supplementary Fig. 9. $P_{final}$ is searched for each parameters set to minimize the σ value. In each panel, values are normalized by those obtained with n = 1. For given n, variations are correlated with color showing the core H content that is calculated with n = 1 and the same impactor $H_2O$ and C abundances, which enables the estimation of model results for any n values based on the results when employing n = 1. Broken and dotted lines show model results as a function of 1/n when 0.2 wt% and 0.6 wt% H are found in the core, respectively.

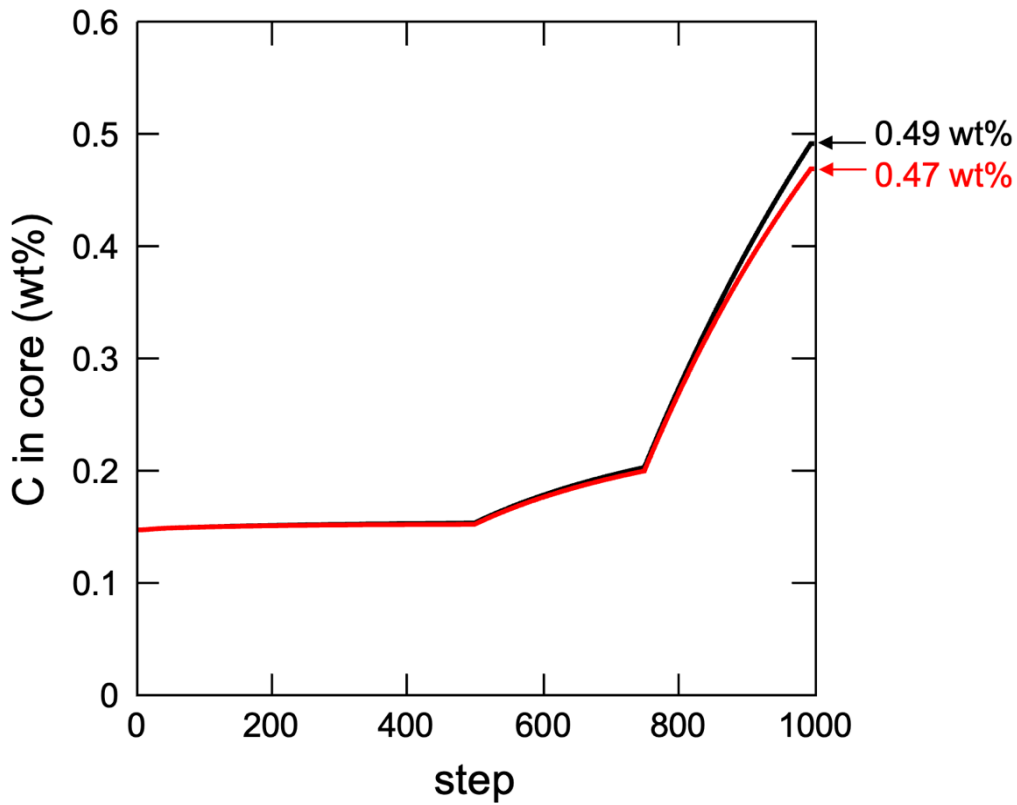

**Supplementary Fig. 12 Effect of C distributions to atmosphere on core C concentration in heterogeneous accretion & multi-stage core formation modelling.** Black line, no carbon in the atmosphere; red line, degassing of all carbon from a magma ocean each step after metal-silicate partitioning (C in the proto-Earth silicate is not inherited to the next step). Parameters were chosen for a typical model shown in Fig. 7: $N = 10$, $n = 1$, $P_{final} = 48$ GPa, 12.5 wt% $H_2O$ and 3.2 wt% C in CI-type materials, no $H_2O$ and C in E-type materials and 25% C loss throughout the accretion. Note that even with $N = 10$ in which the effect of degassing is relatively large (because a larger mass of magma ocean is involved in metal-silicate equilibrium when $N$ is greater), the effect of degassing is minor.

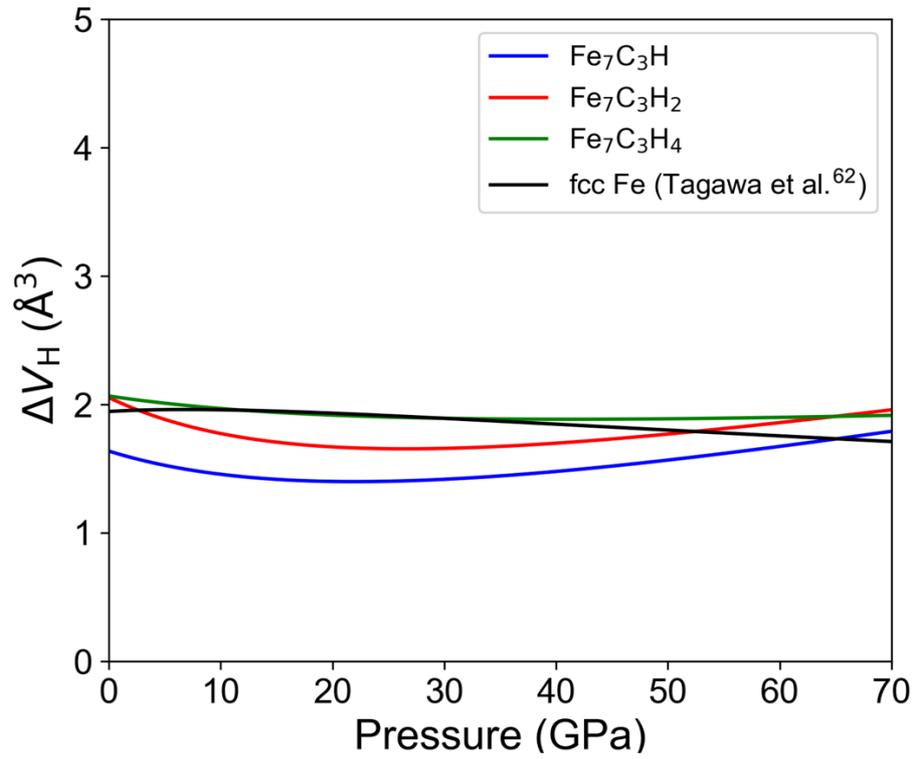

**Supplementary Fig. 13** Comparison of $\Delta V_H$ among $(FeH_x)_7C_3$ (blue, $Fe_7C_3H$; red, $Fe_7C_3H_2$; green, $Fe_7C_3H_4$) and fcc FeH (black)[62]. The equation of state parameters are given in Supplementary Table 5.

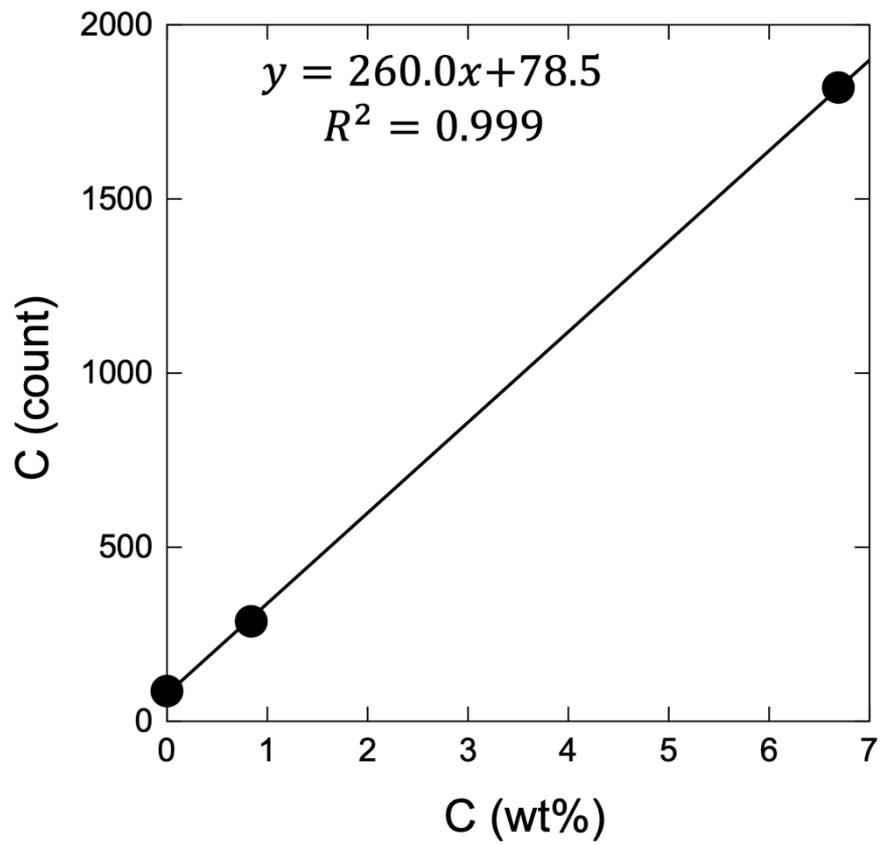

**Supplementary Fig. 14 Calibration curve for the EPMA analyses of carbon based on $Fe_3C$, Fe-0.84wt%C and Re (background).**

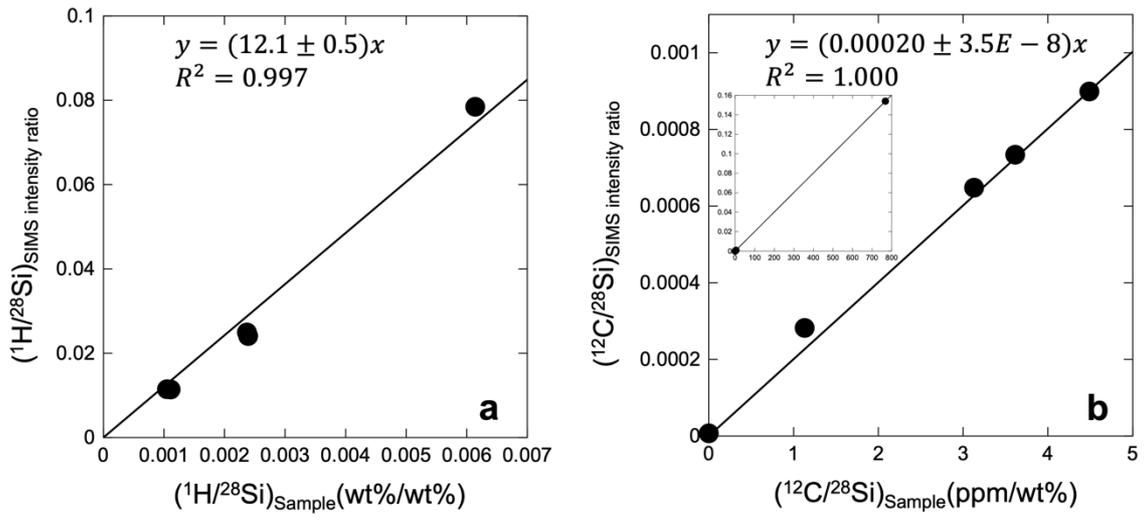

**Supplementary Fig. 15 Calibration curve for SIMS analyses based on standard glasses.**

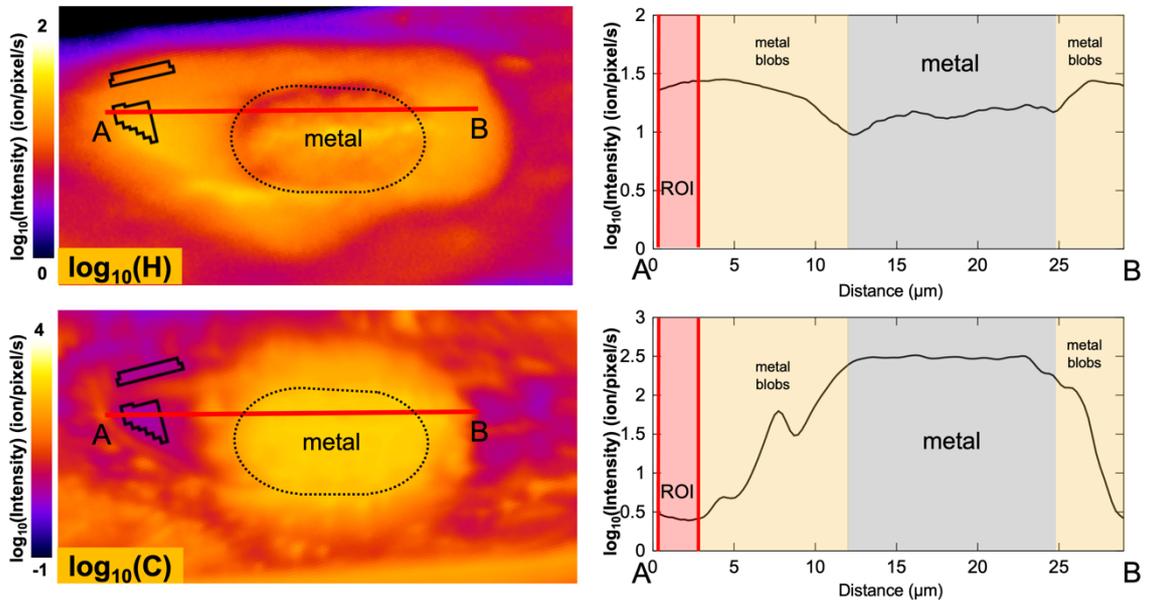

**Supplementary Fig. 16 SCAPS images of secondary ions and the line profiles along red lines A–B.** Same sample as in Fig. 1.